\documentstyle[prb,epsfig,aps,multicol]{revtex}
\def\M#1{$\rm M_{#1}$}
\def\LJ#1{$\rm LJ_{#1}$}
\def\csixty{$\rm C_{60}$}
\def\etal{{\it et al.}}
\begin{document}
\draft
\title{Structural Consequences of the Range of the Interatomic Potential: 
a Menagerie of Clusters}
\author{Jonathan P.~K.~Doye\thanks{Present Address: 
FOM Institute for Atomic and Molecular Physics, 
Kruislaan 407, 1098 SJ Amsterdam, The Netherlands} 
and David J.~Wales}
\address{University Chemical Laboratory, Lensfield Road, Cambridge CB2 1EW, UK}
\date{\today}
\maketitle
\begin{abstract}
We have attempted to find the global minima of clusters containing  between 20 
and 80 atoms bound by the Morse potential as a function of the range of the 
interatomic force.  The effect of decreasing the range is to destabilize strained
structures, and hence the global minimum changes from icosahedral 
to decahedral to face-centred-cubic as the range is decreased.
For $N>45$ the global minima associated with a long-ranged potential have 
polytetrahedral structures involving defects called disclination lines. 
For the larger clusters the network of disclination lines is disordered and the 
global minimum has an amorphous structure resembling a liquid.
The size evolution of polytetrahedral packings enables us to study the 
development of bulk liquid structure in finite systems.  As many experiments 
on the structure of clusters only provide indirect structural information, 
these results will be very useful in aiding the interpretation of experiment.  
They also provide candidate structures for theoretical studies using more 
specific and computationally expensive descriptions of the interatomic 
interactions.  Furthermore, Morse clusters provide a rigorous testing ground for
global optimization methods.
\end{abstract}
\begin{multicols}{2}
\section{Introduction}

Structural information is of fundamental importance in addressing the chemical
and physical properties of any system.
Unfortunately, there is no direct experimental method for determining the 
structure of free clusters in molecular beams. 
Instead, one measures properties which depend on structure and employs models 
of the predicted favoured geometries.
This approach has been combined with
techniques such as electron diffraction,\cite{Farges88}
mass spectral abundances,\cite{Martin96}
chemical reactivity,\cite{Riley94}
magnetism\cite{Bloomfield96}
and x-ray spectroscopy.\cite{Kakar97}
The inversion of the experimental data to obtain structural information, though, 
can be problematic, and always relies on comparisons with the predictions of 
structural models.

Sizes exhibiting special stability are known for certain morphologies and classes
of interatomic potential.\cite{Northby87,JD95d}
The size-dependence of properties sometimes reveals these magic numbers, 
and thus enables a confident structural assignment to be made.  However, 
often rather little is known about the structure between these magic numbers.
One of the most powerful experimental techniques that addresses this deficiency 
is the flow-reactor approach which probes the chemical reactivity 
of size-selected clusters. 
For example, this method has been applied to nickel 
clusters to give detailed information for all sizes up to 
$N$=71.\cite{Parks94,Parks95a,Parks95b,Parks97}
The results show that around $N$=13 and $N$=55 the clusters are icosahedral,
in agreement with the observed magic numbers in other 
experiments.\cite{Bloomfield96,Klots91,Pellarin} 
However, in the size range $29<N<48$ only one structural assignment 
has so far been made
because of the large number of possible geometries to be considered 
and the presence of multiple isomers.\cite{Parks97}

The theoretician can aid in the task of structural assignment by providing
realistic candidate structures.
Indeed, many studies have modelling specific clusters, but 
{\it ab initio} calculations are only feasible for small sizes, 
especially for transition metals, and so empirical potentials are often used.
However, as is clear from the diversity of theoretical results 
obtained for nickel 
clusters,\cite{Stave,Wetzel,Lathiotakis,Montejano,Mei,Nayak,WalesD97b}
consensus between methods is lacking and
it is hard to know which (if any) of the results should be believed.
Even with the simplified description of the interatomic interactions
provided by an empirical potential, it can be an extremely difficult task to 
search the potential energy surface (PES) extensively enough to be confident 
that the global minimum has been found. 
Also, many empirical potentials are too complicated to
provide an understanding of the relationship between the potential and the 
observed structure and so little physical insight is gained.

Therefore, to understand cluster structure there is a need for a hierarchy 
of theoretical models from the general to the specific.
In the present study we use a simple model to understand the structural effects 
of the range of the potential, and so provide one part of the framework
for understanding the physical basis of cluster structure.
We are confident that we have found most of the global minima 
giving the most comprehensive model of cluster structure 
in the small size regime.

\begin{figure}
\begin{center}
\vglue -3mm
\epsfig{figure=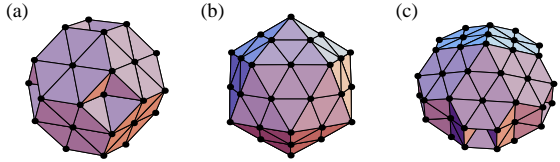,width=8.5cm}
\vglue -3mm
\begin{minipage}{8.5cm}
\caption{(a) 38-atom truncated octahedron, (b) 55-atom Mackay icosahedron, and 
(c) 75-atom Marks decahedron. 
These clusters have optimal shapes for the three main types of ordered packing 
seen in clusters: face-centred cubic (fcc), icosahedral and decahedral, 
respectively. The latter two morphologies cannot be extended to the bulk 
because of the five-fold axes of symmetry.}
\end{minipage}
\label{fig:fid}
\end{center}
\end{figure}

One of the most interesting aspects of cluster structure is the manifestation of 
non-crystallographic symmetries which arise from the absence of translational 
periodicity. 
In particular, many clusters are found to have fivefold axes of symmetry, 
including two of the three main types of ordered structure adopted by 
simple atomic clusters. 
Decahedra have a single fivefold axis of symmetry and are based on
pentagonal bipyramids, while icosahedra have six fivefold axes of symmetry.
The third morphology consists of close-packed clusters. 
Particularly stable examples of each type are illustrated in Fig.\ \ref{fig:fid}.

All the above morpholgies have been observed experimentally. 
Many gas phase clusters have been shown to be icosahedral through the presence of 
the magic numbers associated with the Mackay icosahedra\cite{Mackay} in mass 
spectra: rare gases\cite{Echt81,Harris84,Harris86},
metals,\cite{Pellarin,Martin90,Martin91a,Martin91b}
and molecular clusters.\cite{Echt90,Martin93}
Icosahedral and decahedral structures are also commonly reported for metal 
clusters supported on surfaces,\cite{Marks94}
and more recently fcc and decahedral clusters have been observed for gold 
clusters passivated by 
alkylthiolates.\cite{Whetten96,Andres96,Alvarez97,Cleveland97a,Cleveland97b}

All three structural types are also exhibited by Lennard-Jones (LJ) clusters.
For $N<1600$ icosahedra are most stable; from this size up to $N\approx 10^5$ 
decahedra are most stable and above this fcc clusters.\cite{Raoult89a}
However, these changes do not occur abruptly.
The global minimum of \LJ{38} is the fcc truncated 
octahedron\cite{Pillardy,JD95c} (Fig.\ \ref{fig:fid}(a))
and for at least six sizes with $N<110$ the global minimum is based upon a Marks 
decahedron\cite{JD95d,JD95c} (Fig.\ \ref{fig:fid}(c)).
In this paper we consider a potential with variable range to provide a model 
system which exhibits a much greater diversity of structural behaviour than 
LJ clusters in the small size regime.  
Consequently, the results are relevant to a much wider range of systems.

There have been a number of previous studies on the effect of 
the range of the potential on the structure and phase behaviour of small 
clusters.\cite{JD95c,Braier,Bytheway,JD96a,JD96b,WalesD96,Rey96,Mainz}
These have shown that the number of minima and saddle points
on the PES increases as the range decreases---the PES becomes more 
rugged\cite{Braier,JD96b,HoareM76,HoareM83}---and strained structures are 
destabilized.  The latter effect results in range-induced transitions between 
the ordered morphologies,\cite{JD95c}
and the destabilization and disappearance of the liquid phase as the range is 
decreased.\cite{JD96a,JD96b,Hagen94}
In a previous paper we examined Morse clusters containing up to 
25 atoms and a selection of larger sizes.\cite{JD95c}
Here we consider Morse clusters in the size range $20<N\le80$ atoms. 
Some of the lowest energy structures given here supersede the results of 
the previous paper.

\section{Methods}
\subsection{The potential}
\begin{figure}
\begin{center}
\epsfig{figure=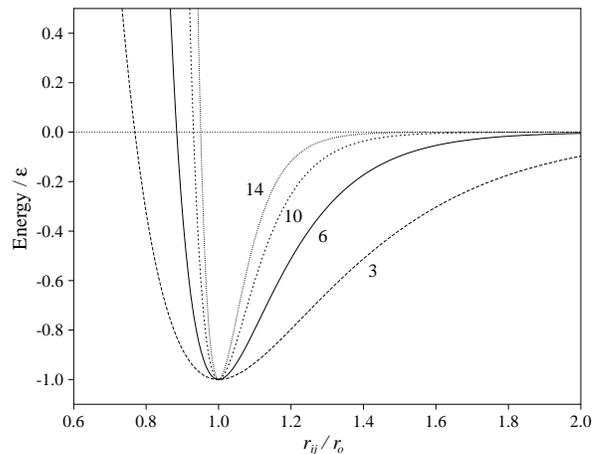,width=8.2cm}
\vspace{3mm}
\begin{minipage}{8.5cm}
\caption{The Morse potential for different values of the range 
parameter $\rho_0$ as indicated.}
\label{fig:Morse}
\end{minipage}
\end{center}
\end{figure}

The Morse potential\cite{Morse} may be written as
\begin{equation}
V_M = \epsilon\sum_{i<j} e^{\rho_0(1-r_{ij}/r_0)}(e^{\rho_0(1-r_{ij}/r_0)}-2),
\end{equation}
where $\epsilon$ is the pair well depth and $r_0$ the equilibrium pair 
separation.  We denote an $N$-atom cluster bound by the Morse potential as M$_N$.
In reduced units ($\epsilon=1$ and $r_0=1$) there is a single adjustable 
parameter, $\rho_0$, which determines the range of the interparticle forces.
Fig.\ \ref{fig:Morse} shows that decreasing $\rho_0$ increases the range of the
attractive part of the potential and softens the repulsive wall, thus widening 
the potential well. 
Values of $\rho_0$ appropriate to a range of materials have been catalogued 
elsewhere.\cite{WalesMD96} The LJ potential has the same curvature at 
the bottom of the well as the Morse potential when $\rho_0=6$.
Girifalco has obtained an intermolecular potential for \csixty\ 
molecules\cite{Girifalco} which is isotropic and short-ranged relative 
to the equilibrium pair separation,
with an effective value of $\rho_0=13.62$.\cite{WalesU94}
The alkali metals have longer-ranged interactions, for example 
$\rho_0$=3.15 has been suggested for sodium.\cite{GirifalcoW}
Fitting to bulk data gives a value of $\rho_0$=3.96 for nickel.\cite{Stave}

At absolute zero the structure with the lowest free energy is simply the global 
potential energy minimum of the PES.
At higher temperatures, entropic factors must also be considered. 
Although we only perform a comprehensive survey of the zero Kelvin geometries, 
the structural effects of temperature are considered in the subsequent 
discussion.

To understand the structural effects of the range parameter,
$\rho_0$, it is instructive to look more closely at the form of the potential.
The energy can be partitioned into three contributions:
\begin{equation}
V_M=-n_{nn}\epsilon+E_{\rm strain}+E_{nnn}.
\end{equation}
The number of nearest-neighbour contacts, $n_{nn}$, the strain energy, 
$E_{\rm strain}$, and the contribution to the energy from non-nearest neighbours,
$E_{nnn}$, are given by
\begin{eqnarray}
n_{nn}&=&\sum_{i<j, x_{ij}<x_0}1, \nonumber \\
E_{\rm strain}&=&\epsilon\sum_{i<j, x_{ij}<x_0}
                (e^{-\rho_0 x_{ij}}-1)^2, \nonumber \\
E_{nnn}&=&\epsilon\sum_{i<j, x_{ij}>x_0}e^{-\rho_0 x_{ij}}(e^{-\rho_0 x_{ij}}-2),
\end{eqnarray}
where $x_{ij}=r_{ij}/r_0-1$, and $x_0$ is a nearest-neighbour criterion.
$x_{ij}$ is the strain in the contact between atoms $i$ and $j$.

The dominant term in the energy comes from $n_{nn}$. $E_{nnn}$ is a smaller 
term and its value varies in a similar manner to $n_{nn}$. 
It is only likely to be important in determining the lowest energy 
structures when other factors are equal. 
For example, bulk fcc and hexagonal close-packed (hcp) lattices both have 
twelve nearest neighbours per atom.
Next-nearest neighbour interactions are the cause of the lower energy of the 
hcp crystal when a pair potential such as the LJ form is used.\cite{LJ,Kihara}

$E_{\rm strain}$, which measures the energetic penalty for the deviation of a 
nearest-neighbour distance from the equilibrium pair distance, is a key 
quantity in our analysis.
It must not be confused with strain due to an applied external force.
For a given geometry, $E_{\rm strain}$ grows rapidly with increasing $\rho_0$
because the potential well narrows.
To a first approximation the strain energy grows quadratically with 
$\rho_0$.\cite{JD95c} Hence, decreasing the range destabilizes strained 
structures.

From the above analysis we can see that minimization of the potential energy 
involves a balance between maximizing $n_{nn}$ and minimizing $E_{\rm strain}$. 
The interior atoms of the three morphologies (Fig.\ \ref{fig:fid}) 
all have twelve nearest neighbours, and
so differences in $n_{nn}$ arise from surface effects.
The optimal shape for each morphology results from the balance between 
maximizing the proportion of $\{111\}$ faces (an atom in a $\{111\}$ face 
is 9-coordinate, but in a $\{100\}$ face only 8-coordinate) and minimizing 
the fraction of atoms in the surface layer.
As Mackay icosahedra (Fig.\ \ref{fig:fid}(b)) have only $\{111\}$ faces and 
are approximately spherical, the icosahedra have the largest $n_{nn}$. 
Complete Mackay icosahedra occur at $N=13, 55, 147, \ldots$
A pentagonal bipyramid has only $\{111\}$ faces, 
but because it is not very spherical 
more stable decahedral forms are obtained by truncating the structure parallel 
to the five-fold axis to reveal five $\{100\}$ faces and then
introducing re-entrant $\{111\}$ faces between adjacent $\{100\}$ faces.
The resulting structure is called a Marks decahedron (Fig.\ \ref{fig:fid}(c)) and
was first predicted by the use of a modified Wulff construction.\cite{Marks84}
Decahedra generally have lower values of $n_{nn}$ than icosahedra 
because of the $\{100\}$ faces.
The tetrahedron and octahedron are fcc structures that have only $\{111\}$ 
faces, but they are not very spherical. 
The optimal fcc structure is the truncated octahedron with regular hexagonal 
faces (Fig.\ \ref{fig:fid}(a)). 
At larger sizes than those considered in this study, the optimal structure 
involves further facetting, 
so that it more closely approximates the Wulff polyhedron.\cite{Cleveland91}
Of the three morphologies fcc structures have the smallest values of $n_{nn}$.
\begin{figure}
\begin{center}
\vglue -5mm 
\epsfig{figure=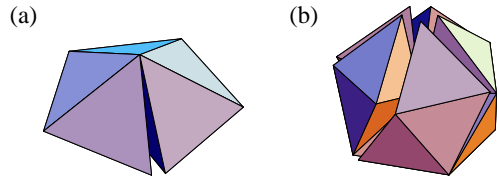,width=8.2cm}
\vglue -5mm 
\begin{minipage}{8.5cm}
\caption{Examples of the strain involved in packing tetrahedra. 
(a) Five regular tetrahedra around a common edge produce a gap of $7.36^\circ$.
(b) Twenty regular tetrahedra about a common vertex
produce gaps equivalent to 1.54 steradians.}
\end{minipage}
\label{fig:gaps}
\end{center}
\end{figure}

A Mackay icosahedron can be decomposed into twenty fcc tetrahedra, but as we 
see from Fig.\ \ref{fig:gaps}(b), when twenty regular tetrahedra are packed 
around a common vertex large gaps remain. 
To bridge these gaps the tetrahedra have to be distorted, thus introducing 
strain.  The distance between adjacent vertices of the icosahedron is 
5\% larger than the distance between a vertex and the centre. 
Similarly, a pentagonal bipyramid can be decomposed into five fcc tetrahedra 
sharing a common edge. Again, a gap remains if regular tetrahedra are used 
(Fig.\ \ref{fig:gaps}(a)) and consequently decahedral
structures are strained, although not as much as icosahedra.
In contrast, close-packed structures can be unstrained.

Having deduced the relative values of $n_{nn}$ and $E_{\rm strain}$ for 
icosahedral, decahedral and fcc structures, 
we can predict the effect of the range on the competition between them.
For a moderately long-ranged potential, the strain associated with the 
icosahedron can be accommodated without too large an energetic penalty 
and so such structures are the most stable.
As the range of the potential is decreased, the strain energy associated 
with icosahedra
increases rapidly, and there comes a point where decahedra become more stable. 
Similarly, for a sufficiently short-ranged potential fcc structures become 
more stable than decahedral structures.

The above decomposition of the potential energy also helps us to understand the
effect of size on the energetic competition between the three morphologies. 
The differences in $n_{nn}$, which arise from the different 
surface structures, are approximately proportional to the surface area 
($\propto N^{2/3}$). 
The strain energies, however, are proportional to the volume ($\propto N$). 
Therefore, the differences in $E_{\rm strain}$ increase more rapidly with 
size than the differences in $n_{nn}$, thus explaining the change in the most 
stable morphology from icosahedral to decahedral to fcc with increasing size.
The effect of increasing the size is similar to the effect of decreasing 
the range of the potential: both destabilize strained structures. 

\subsection{Searching the potential energy surface}The principal method that 
we have used to generate candidate structures for the global minima makes 
use of the physical insight gained from the last section. 
We have simply attempted to construct geometries that maximize $n_{nn}$ 
for the three ordered morphologies.\cite{JD95d}
The resulting structures were then optimized by either conjugate 
gradient\cite{Recipes} or eigenvector-following\cite{Cerjan} methods.
A similar approach was successfully used by Northby to generate 
low energy icosahedra for LJ clusters.\cite{Northby87}
The effectiveness of this method is demonstrated by how few of Northby's 
lowest energy structures have been superseded 
and by the length of time that it has taken to find these 
exceptions.\cite{JD95d,Pillardy,JD95c,Coleman,Xue,Deaven96,WalesD97}
Our approach, however, depends on the imagination of the practitioner 
to conceive of all the possible ways that a structure with a large value of 
$n_{nn}$ could be obtained.
Furthermore, this method will always fail to find the global minimum if the 
latter is not based on one of the ordered structures, as is the case for the 
larger clusters we have considered at low values of $\rho_0$. 

To complement the above approach two global optimization techniques were used 
to try to find structures that might have been missed.
Firstly, we used a method in which eigenvector-following 
is employed to take steps directly between minima on the PES.\cite{JD97a,Barkema}
If low temperature Metropolis Monte Carlo is used in this space of minima, 
the system will walk down to the bottom of a basin containing many minima.
This technique avoids the difficulties associated with trapping in local minima 
that can occur for methods which take steps directly in configuration space. 
Secondly, we used 
a `basin hopping' approach, which has proved to be effective
for LJ clusters;\cite{WalesD97,JD97d} 
it is the only unbiased global optimization
method to have found the global minima that are Marks decahedra.
For Morse clusters it was able to find all the lowest energy minima at 
$\rho_0$=3, 6, 10 and all but twelve at $\rho_0$=14; this included some 
structures that were lower than any of those we had constructed.
The fact that we found most of the minima both by unbiased global optimization 
and by construction makes us confident that our lowest energy structures 
are truly global minima.

The results for the basin-hopping algorithm are impressive because
the global optimization task for Morse clusters is a difficult one. 
The size of the configuration space for the larger clusters 
is compounded for short-ranged potentials by the nature of the PES.
Firstly, the PES becomes more rugged as the range is decreased.\cite{JD96b}
The physical reason for the larger number of minima at short range is
the loss of accessible configuration space as the potential wells become 
narrower, producing barriers where there are none at long range.
Stillinger and Stillinger\cite{StillD90b} and Bytheway and Kepert\cite{Bytheway} 
both found that minimizations performed from random starting configurations 
are much less likely to find the global minimum for a short-ranged potential.

Similarly, barrier heights are also likely to become higher and 
rearrangements more localized for a shorter-ranged potential.
These trends have been observed in comparisons of the rearrangements of 
55-particle \csixty\ and LJ clusters.\cite{Wales94b}
Hence the range of the potential is likely to have a significant impact
on the dynamics, making escape from local minima much more difficult. 
This effect has been observed by Rose and Berry who have shown that the rate 
at which the ground state structure of a potassium chloride cluster is found 
upon cooling can be significantly decreased by using a shielded Coulomb 
potential to reduce the range of the interactions.\cite{Rose93b}

Furthermore, as a result of the competition between the decahedral 
and the various types of close-packed structures for short-ranged potentials,
it is more likely that there are a number of low energy minima which are 
very close in energy but are structurally dissimilar. 
Each of these minima lie at the bottom of their own funnel on the PES.
This multiple funnel topography can lead to cases where optimization 
is extremely difficult because the free energy barriers for 
transitions between the funnels can be large, thus leading to trapping. 
The worst cases are when relaxation down the PES preferentially takes
the system into a funnel which does not end at the global minimum,\cite{JD96c}
and when the global minimum only becomes the state with the lowest free energy 
at low temperatures.\cite{JD97d}

Finally, fcc and decahedral minima are more structurally dissimilar from
minima typical of the liquid-like state than the icosahedral structures. 
Therefore, the paths from the liquid to these structures are likely to 
be fewer and longer than those leading to the icosahedral structures,\cite{JD97a} 
and so relaxation down the PES from the liquid-like state to fcc and decahedral 
global minima is harder.

For a number of reasons Morse clusters are an ideal system with which to test 
global optimization methods designed for configurational problems. 
Firstly, in this paper we provide very good estimates for the energies of the 
global minima.
Secondly, this system represents a much more general---the global minima have a 
variety of structural types---and 
tougher examination than is provided by LJ clusters, 
a much-used test system for global optimization algorithms.\cite{WalesD97}
And finally, the results for our basin-hopping algorithm provide a benchmark 
that any would-be global optimization method should try to beat. 
\end{multicols}
\begin{figure}
\begin{center}
\epsfig{figure=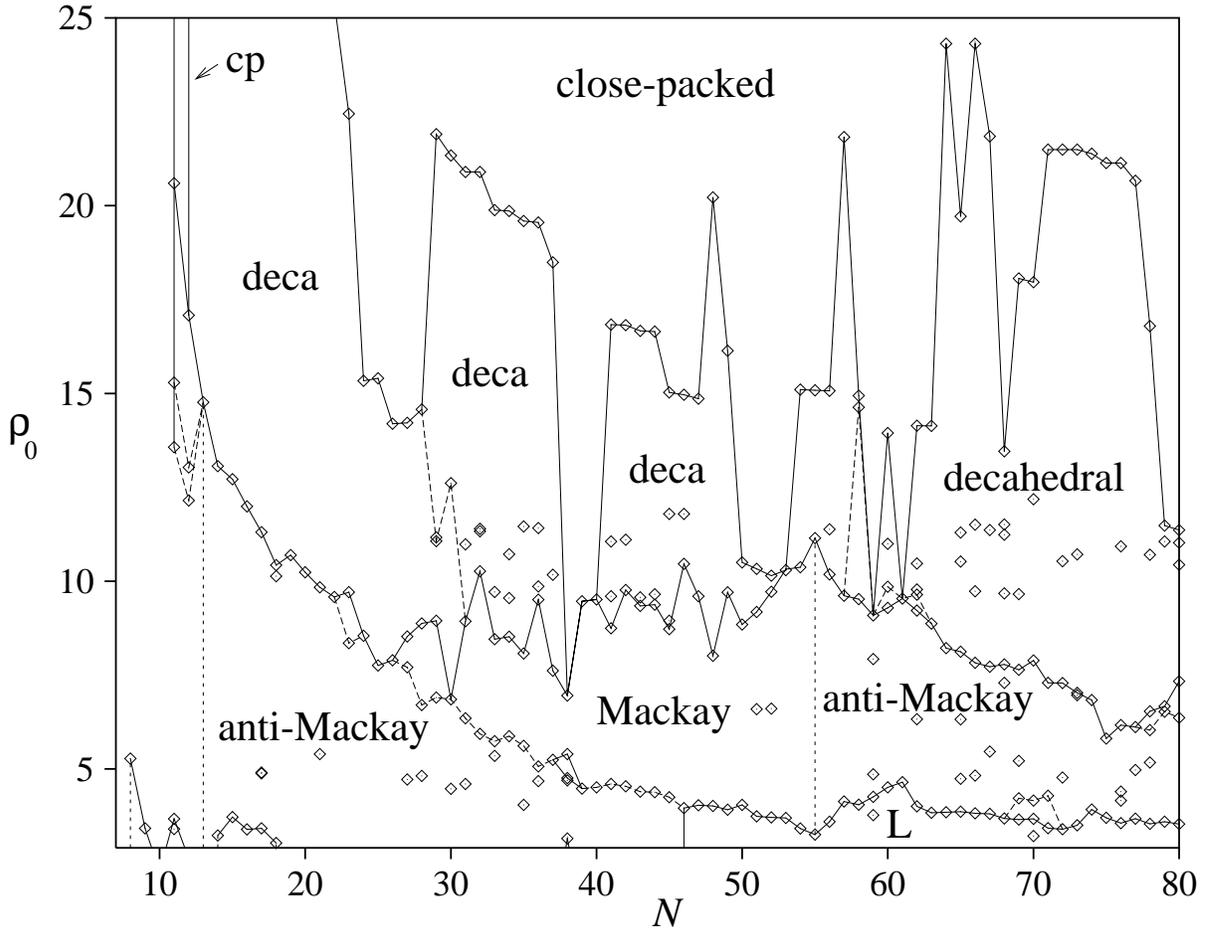,width=17.2cm}
\vspace{3mm}
\caption{Zero temperature `phase diagram' showing the variation of the lowest 
energy structure with $N$ and $\rho_0$.
The data points are the values of $\rho_0$ at which the global minimum changes.
The lines joining the data points divide the phase diagram into regions where 
the global minima have similar structures. 
The solid lines denote the boundaries between the four main structural 
types---icosahedral, decahedral, close-packed and those associated with low 
$\rho_0$ (L)---and the dashed lines are internal 
boundaries within a structural type, e.g.~between icosahedra with Mackay and 
anti-Mackay overlayers, or between decahedra with different length decahedral 
axes.}
\label{fig:phase}
\end{center}
\end{figure}
\begin{multicols}{2}
\section{Results}All the global minima that we have found are catalogued in 
Table 1 along with their energies, point groups, number of nearest neighbours, 
strain energies and the values of $\rho_0$ for which they are probably the 
lowest energy minimum.

The results are summarized in Fig.\ \ref{fig:phase} which provides a 
zero temperature `phase diagram', 
showing how the global minimum depends upon $N$ and $\rho_0$. 
The structural behaviour of Morse clusters with fewer than eight atoms is 
rather uninteresting because the global minimum is independent of $\rho_0$.
For all $N\ge 8$, however, the global minimum changes at least once as a 
function of $\rho_0$.
For $N\ge 13$, icosahedral, decahedral and fcc structures all exist, 
and the form of the phase diagram is in good agreement with the predictions we 
made earlier based upon the decomposition of the potential energy. 
For most sizes the structure changes from icosahedral to decahedral to 
close-packed as the range of the potential is decreased. 
For $N<23$, however, the transition from decahedral to close-packed occurs 
at larger values of $\rho_0$ than we consider in this study. 
There are also a number of sizes ($N$=38--40, 52, 53, 59 and 61) for which there
is a transition directly from an icosahedral to a close-packed structure; 
this occurs when $n_{nn}$ for the lowest energy close-packed structure is 
greater than or equal to that for the lowest energy decahedron.
\begin{figure}
\begin{center}
\epsfig{figure=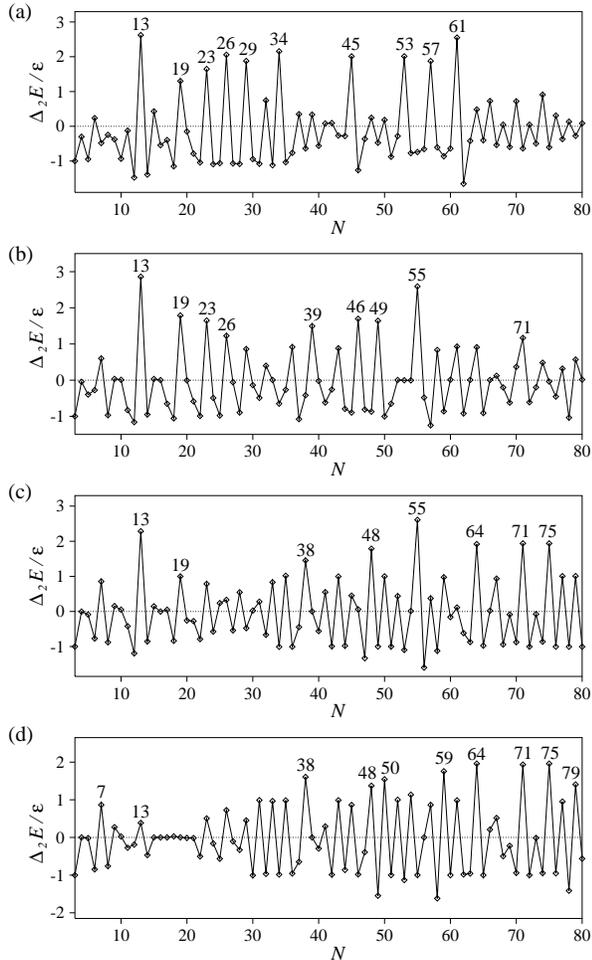,width=8.2cm}
\vspace{3mm}
\begin{minipage}{8.5cm}
\caption{Plots of $\Delta_2 E$ as a function of $N$ for 
(a) $\rho_0$=3, (b) $\rho_0$=6, (c) $\rho_0$=10 and (d) $\rho_0$=14.} 
\end{minipage}
\label{fig:D2E}
\end{center}
\end{figure}

The boundaries between the different morphologies are sensitive functions of $N$. 
Such size dependence is observed for many properties of clusters, and gradually 
lessens as $N$ increases (because the addition of a single atom becomes a smaller 
perturbation) until the bulk limit is reached.
The decahedral to close-packed boundary is particularly sensitive, 
because the range of $\rho_0$ for which the decahedron is most stable 
changes dramatically even when the difference in $n_{nn}$ between the 
decahedral and close-packed structures changes by one.

\begin{figure}
\begin{center}
\epsfig{figure=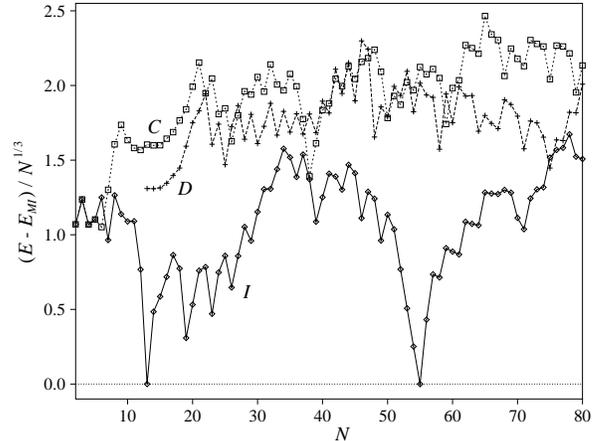,width=8.2cm}
\vspace{3mm}
\begin{minipage}{8.5cm}
\caption{Comparison of the energies of icosahedral (solid line with diamonds), 
decahedral (dashed line with crosses) and close-packed 
(dotted line with squares) \M N clusters at $\rho_0$=6.
The energy zero is $E_{MI}$, the interpolated energy of Mackay icosahedra. 
$E_{MI}=-3.0354+0.2624 N^{1/3}+8.8400 N^{2/3}-6.8381N$ and was obtained by 
fitting to the first four Mackay icosahedra ($N$=13, 55, 147 and 309).}
\end{minipage}
\label{fig:comp}
\end{center}
\end{figure}

Sizes for which a morphology is the lowest in energy for a particularly large 
range of $\rho_0$ indicate that the structure is especially stable. 
The optimal geometries shown in Fig.\ \ref{fig:fid} are good examples.
Another indicator of special stability is provided by 
$\Delta_2 E(N)=E(N+1)+E(N-1)-2 E(N)$.
Peaks in $\Delta_2E$ have been found to correlate well with the magic numbers 
(sizes at which clusters are particularly abundant) observed 
in mass spectra.\cite{Clemenger}
Plots of $\Delta_2 E$ are shown in Fig.\ \ref{fig:D2E} for a number of values 
of $\rho_0$. 
Unsurprisingly, the plot for $\rho_0$=6 is very similar to that for LJ clusters 
with peaks due to especially stable icosahedral clusters.
At higher values of $\rho_0$ peaks corresponding to close-packed 
and decahedral clusters begin to occur. 
The plot at $\rho_0$=14 is very similar to that recently obtained for \csixty\ 
clusters using the Girifalco intermolecular potential.\cite{JD96d}
If the energy is \lq normalized' by subtracting a suitable function of $N$, 
particularly stable sizes again stand out (Fig.\ \ref{fig:comp}).

In the following subsections we will look at the growth sequences for 
each morphology in more detail. We also examine the unusual structures
that occur for the larger clusters at low $\rho_0$, which, as we will see, 
involve a mixture of order and disorder.

\subsection{Icosahedral clusters}
\begin{figure}
\begin{center}
\epsfig{figure=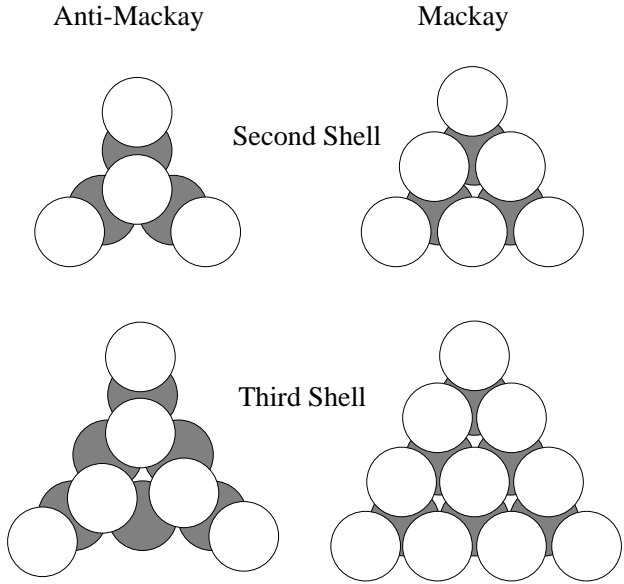,width=8.2cm}
\vglue 3mm 
\begin{minipage}{8.5cm}
\caption{Atomic positions for the two possible overlayers of the icosahedron, 
anti-Mackay (left) and Mackay (right).
These are shown for a single face of the icosahedron.}
\end{minipage}
\label{fig:over}
\end{center}
\end{figure}

Many small clusters are polytetrahedral 
in the sense that the whole of the cluster can be divided into tetrahedra.
This category includes the 13-atom icosahedron, which can be decomposed into
twenty tetrahedra sharing a common vertex.  
Addition of atoms to the icosahedron can occur in two ways
and the two types of overlayer that result are illustrated 
in Fig.\ \ref{fig:over}.
One growth mode (fcc-like) continues the fcc packing of the twenty strained 
tetrahedra making up the icosahedron, and leads to the 
55-atom Mackay icosahedron (Fig.\ \ref{fig:fid}(b)).
This scheme introduces octahedral interstices, and so the resulting 
structures are no longer polytetrahedral.
The other \lq anti-Mackay' (hcp-like) growth mode involves sites which are 
hcp with respect to the tetrahedra.
For growth on the 13-atom icosahedron, this overlayer 
preserves polytetrahedral character. 
Each of the vertices of the original icosahedron becomes icosahedrally 
coordinated, and the structure that results from the completion of this 
overlayer, 45A, is a rhombic tricontahedron; 
it is an icosahedron of interpenetrating icosahedra.
Interestingly, the rhombic tricontahedron is the face-dual of the 
truncated icosahedron made famous by \csixty; 
indeed, it is even a particularly stable shell for the 
coverage of a \csixty\ molecule by alkaline earth metal atoms.\cite{Martin94b}
In previous studies, the anti-Mackay overlayer has been referred to as the 
polyicosahedral\cite{Farges88} or the face-capping overlayer.\cite{Northby87}
Such names are reasonable for growth on the 13-atom icosahedron, but are 
confusing for growth on larger Mackay icosahedra.

\begin{figure}
\vglue -0.5mm
\begin{center}
\epsfig{figure=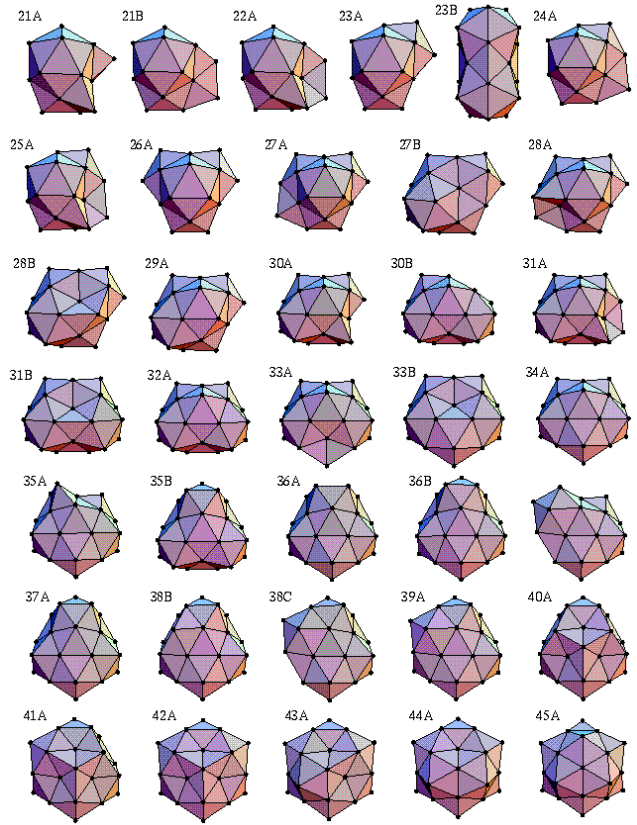,width=8.5cm}
\vglue -0.2mm
\begin{minipage}{8.5cm}
\caption{Icosahedral global minima formed by growth of an anti-Mackay overlayer 
on the 13-atom icosahedron, except structure 23B which is composed of 
two face-sharing icosahedra.}
\end{minipage}
\label{fig:icos.am}
\end{center}
\end{figure}

\begin{figure}
\begin{center}
\epsfig{figure=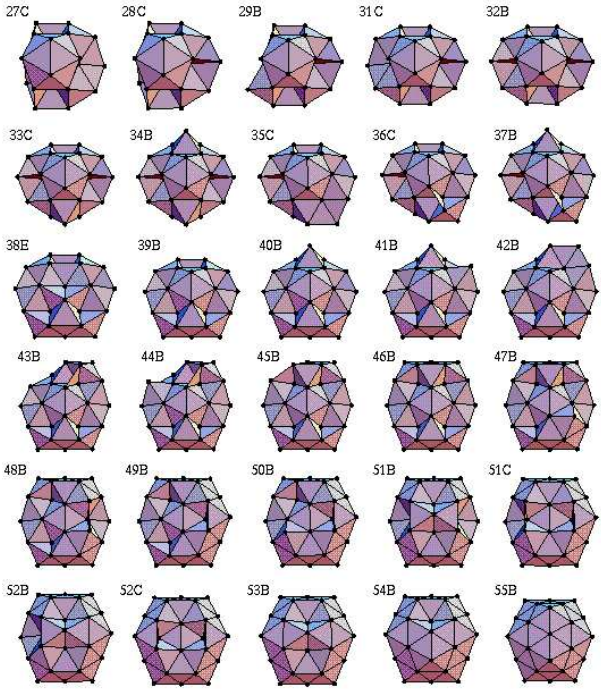,width=8.5cm}
\begin{minipage}{8.5cm}
\caption{Icosahedral global minima formed by growth of a Mackay overlayer on 
the 13-atom icosahedron.}
\label{fig:icos.M}
\end{minipage}
\end{center}
\end{figure}

The icosahedral structures with an anti-Mackay overlayer are illustrated in 
Fig.\ \ref{fig:icos.am} and those with a Mackay overlayer 
in Fig.\ \ref{fig:icos.M}.
Growth from the 13-atom icosahedron begins in the anti-Mackay
positions, because these do not include any low-coordinate edge sites, 
thus giving a larger $n_{nn}$.
However, at some critical size the Mackay overlayer becomes lower in energy 
because of the larger strain energies associated with the anti-Mackay overlayer.
Further growth then leads to the next Mackay icosahedron.
The size at which this change occurs depends on the range of the potential;
it is given by the dashed line of negative slope that divides the 
icosahedral region of the phase diagram (Fig.\ \ref{fig:phase}).
The crossover size increases with the range of the potential.

At $\rho_0=6$ the Mackay overlayer is the lowest in energy for $N\ge32$; 
the corresponding result for LJ clusters is $N\ge31$.
In {\it ab initio} molecular dynamics calculations for lithium 
clusters,\cite{Sung} the anti-Mackay overlayer is lowest in energy up to $N=45$. 
Similarly, polytetrahedral structures were observed in a tight-binding study 
of sodium clusters up to the largest size considered $N$=34 
(for which 34A was the most stable structure).\cite{Poteau}
These effects are nicely explained by the long range of alkali metal potentials.

Especially stable structures with an anti-Mackay overlayer occur when the 
icosahedral coordination of a vertex atom is complete. 
These structures give rise to the peaks in $\Delta_2 E$ at 
$N$=19, 23 26, 29, 34 and 45 for $\rho_0$=3; 
only the first four of these peaks are seen at $\rho_0$=6, 
and only the first at $\rho_0$=10 (Fig.\ \ref{fig:D2E}).
Some of these magic numbers have been observed in the mass spectra
of noble gases\cite{Echt81,Harris84,Harris86} and even barium.\cite{Rayane}
The centres of the icosahedra in these structures form a dimer for 19A, 
an equilateral triangle for 23A, a tetrahedron for 26A, 
a trigonal bipyramid for 29A, a pentagonal bipyramid for 34A, 
and an icosahedron for 45A.

At a number of sizes there is more than one global minimum with 
an anti-Mackay overlayer for different values of $\rho_0$.
The transitions between these structures are related to small differences in the 
value of $E_{\rm strain}$.  
Also illustrated in Fig.\ \ref{fig:icos.am} is structure 23B, 
which is made up of two face-sharing icosahedra. 
It can be formed from structure 17B by the addition of six atoms 
to part of the overlayer. 

The first global minimum with a Mackay overlayer occurs at $N$=27. 
Especially stable structures occur at those sizes for which complete faces of 
the 55-atom Mackay icosahedra are missing. 
These structures give rise to the peaks at $N$=39, 46 and 49 for $\rho_0$=6 and
correspond to 5, 2 and 1 missing faces, respectively. Again, these magic numbers 
have been seen in the mass spectra of noble gases.\cite{Harris84}

Structure 38E has an atom missing from a vertex of the original 13-atom 
icosahedron to allow the overlayer to complete a particularly stable form.
This structure was not found in Northby's study,
but it is the lowest energy LJ icosahedral cluster.\cite{Deaven96}

\begin{figure}
\begin{center}
\epsfig{figure=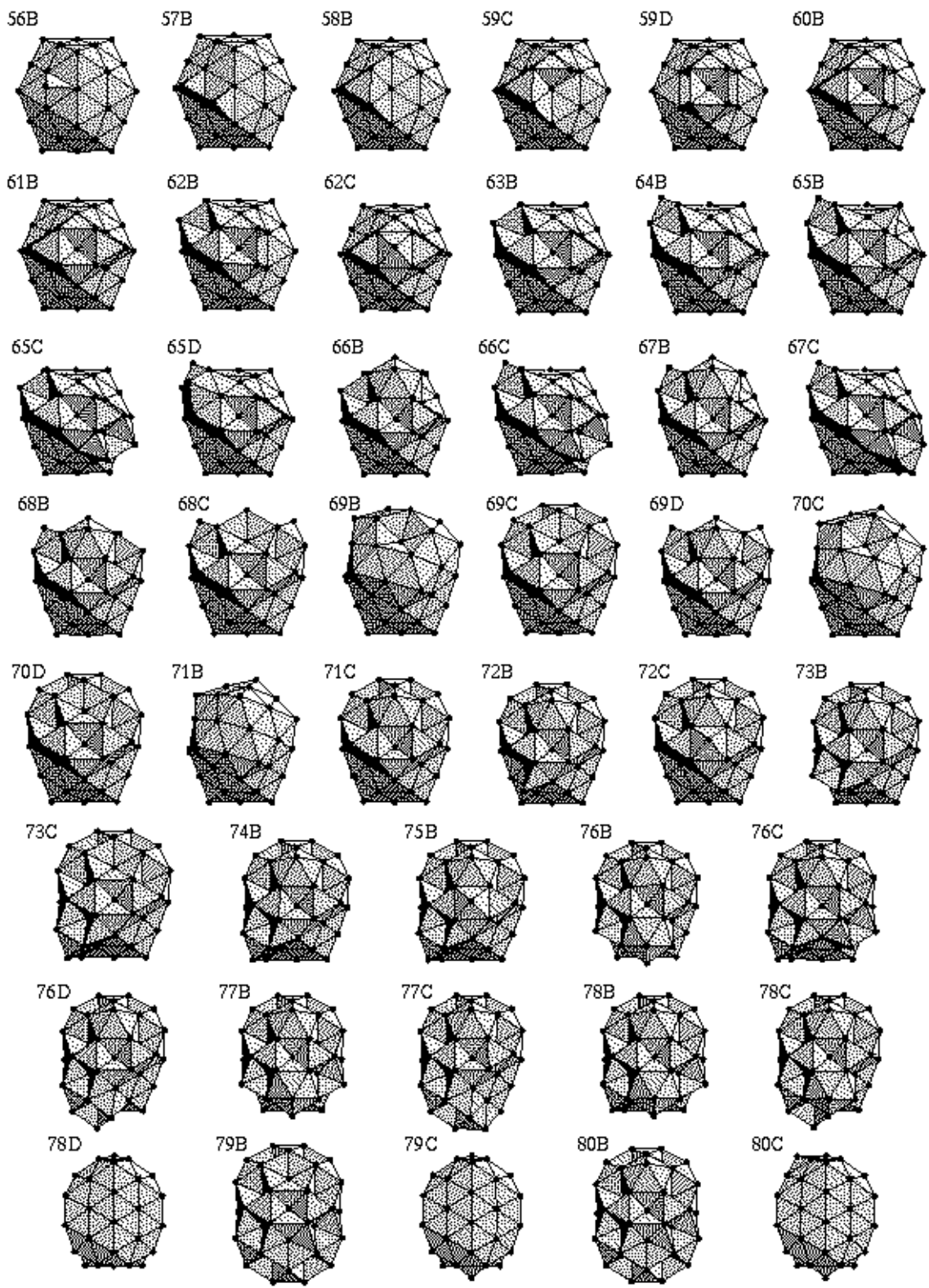,width=16.0cm}
\caption{Icosahedral global minima formed by growth from the 55-atom Mackay 
icosahedron.}
\label{fig:icos.55+}
\end{center}
\end{figure}

The icosahedral global minima with more than 55 atoms are 
shown in Fig.\ \ref{fig:icos.55+}.
As for the 13-atom icosahedron, growth initially occurs at the anti-Mackay 
sites, because this results in structures with larger $n_{nn}$. 
Completion of this overlayer occurs for a cluster with 127 atoms.
The vertices of the 55-atom Mackay icosahedron become icosahedrally coordinated,
but the edge atoms have a decahedral coordination shell 
(this is clearly visible for 59D) leading to the half octahedra that are 
visible in the surface layers of the anti-Mackay clusters.
The most prominent peak in $\Delta_2E$ due to an anti-Mackay 
structure occurs at $N$=71 for $\rho_0$=6. 
This corresponds to an overlayer which completely covers the five faces 
surrounding a vertex of the underlying icosahedron. 
There are smaller peaks at $N$=58, 61, and 64, which 
correspond to complete coverage of one, two and three faces, respectively.

Structure 69C has a vertex atom missing from the underlying 
Mackay icosahedron like 38A. 
There are also structures (62C, 65D, 72C and 75B) where an atom is added to 
the surface of the overlayer rather than to the Mackay icosahedron. 
Again there are transitions between different anti-Mackay structures
resulting from small differences in $E_{\rm strain}$.

The first structure in this size range with a Mackay overlayer is 78D. 
We expect that the crossover from an anti-Mackay to a Mackay overlayer
will again shift to larger size as the range increases, 
but we have not investigated this prediction in the present work.  
There are also three icosahedral global minima, 69B, 70C and 71B, 
which do not fit neatly into either the Mackay or anti-Mackay category. 
Their surface layers have a Mackay-like character, but are not in correct 
alignment with the underlying icosahedron. 
The overlayer has been given a twist about one of the fivefold axis in order to 
collapse some of the half octahedra at the edge of the overlayer into trigonal 
bipyramids in a multiple diamond-square-diamond (DSD) process.\cite{Lipscomb}
In fact for \M{71} at $\rho_0$=5, the lowest energy Mackay structure is 
a $C_{5v}$ transition state corresponding to a multiple DSD rearrangement 
between two permutational isomers of 71B.

Fig.\ \ref{fig:comp} illustrates the variation of the icosahedral 
energies with size.
The complete Mackay icosahedra appear as narrow minima separated by broad 
maxima corresponding to structures with incomplete outer shells. 
At $\rho_0$=6 it is only near the top of these maxima that fcc and decahedral 
structures begin to have similar energies, for example at $N$=38 and 75.

\subsection{Decahedral clusters}
\begin{figure}
\begin{center}
\epsfig{figure=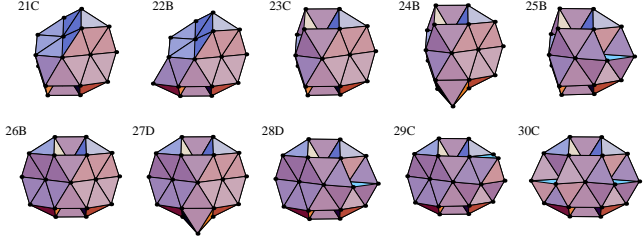,width=9.0cm}
\begin{minipage}{8.5cm}
\caption{Global minima based upon a decahedron with three atoms along the 
fivefold axis.}
\end{minipage}
\label{fig:deca3}
\end{center}
\end{figure}

\begin{figure}
\begin{center}
\epsfig{figure=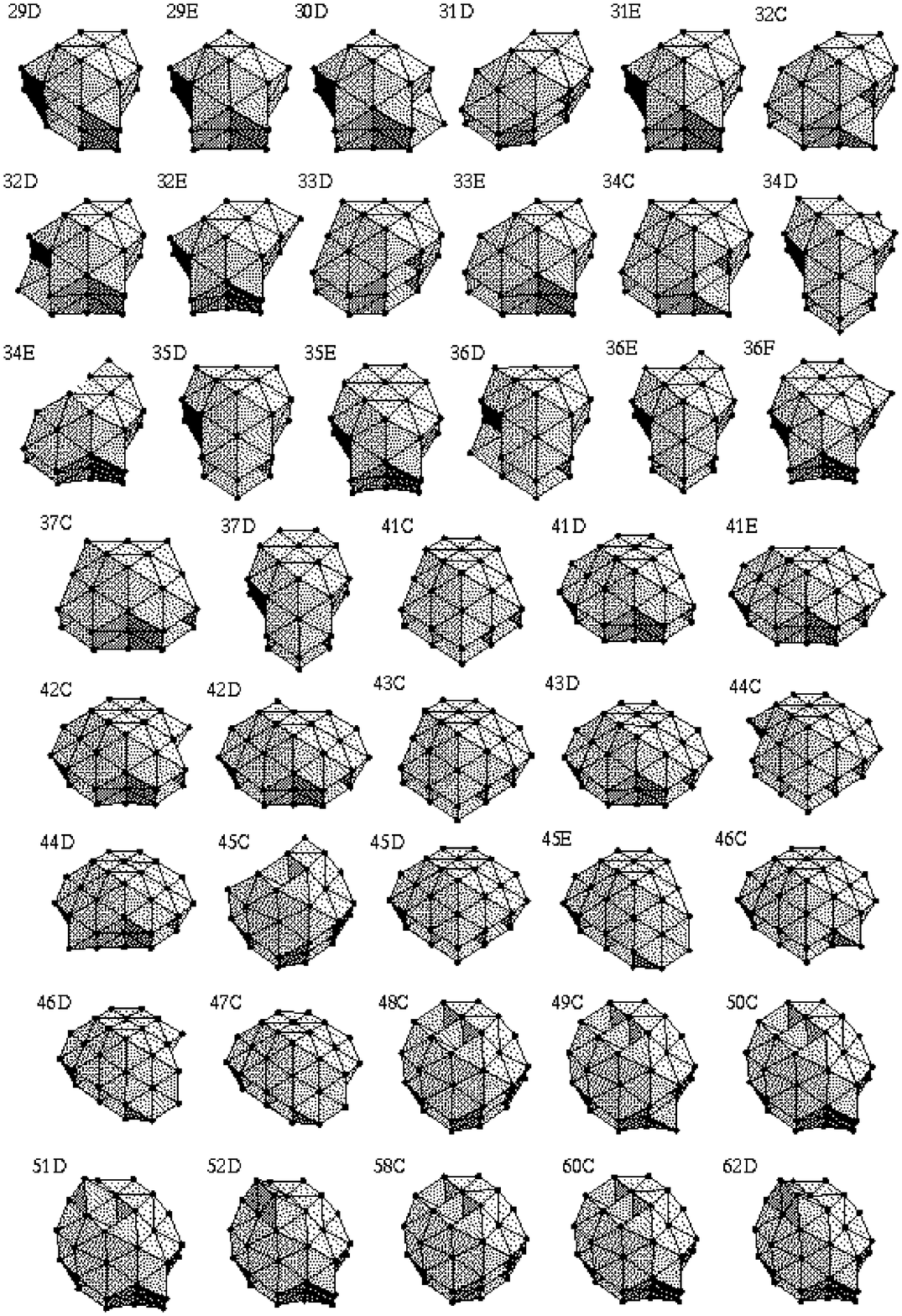,width=15.5cm}
\caption{Global minima based upon a decahedron with four atoms along 
the fivefold axis.}
\label{fig:deca4}
\end{center}
\end{figure}

\begin{figure}
\begin{center}
\epsfig{figure=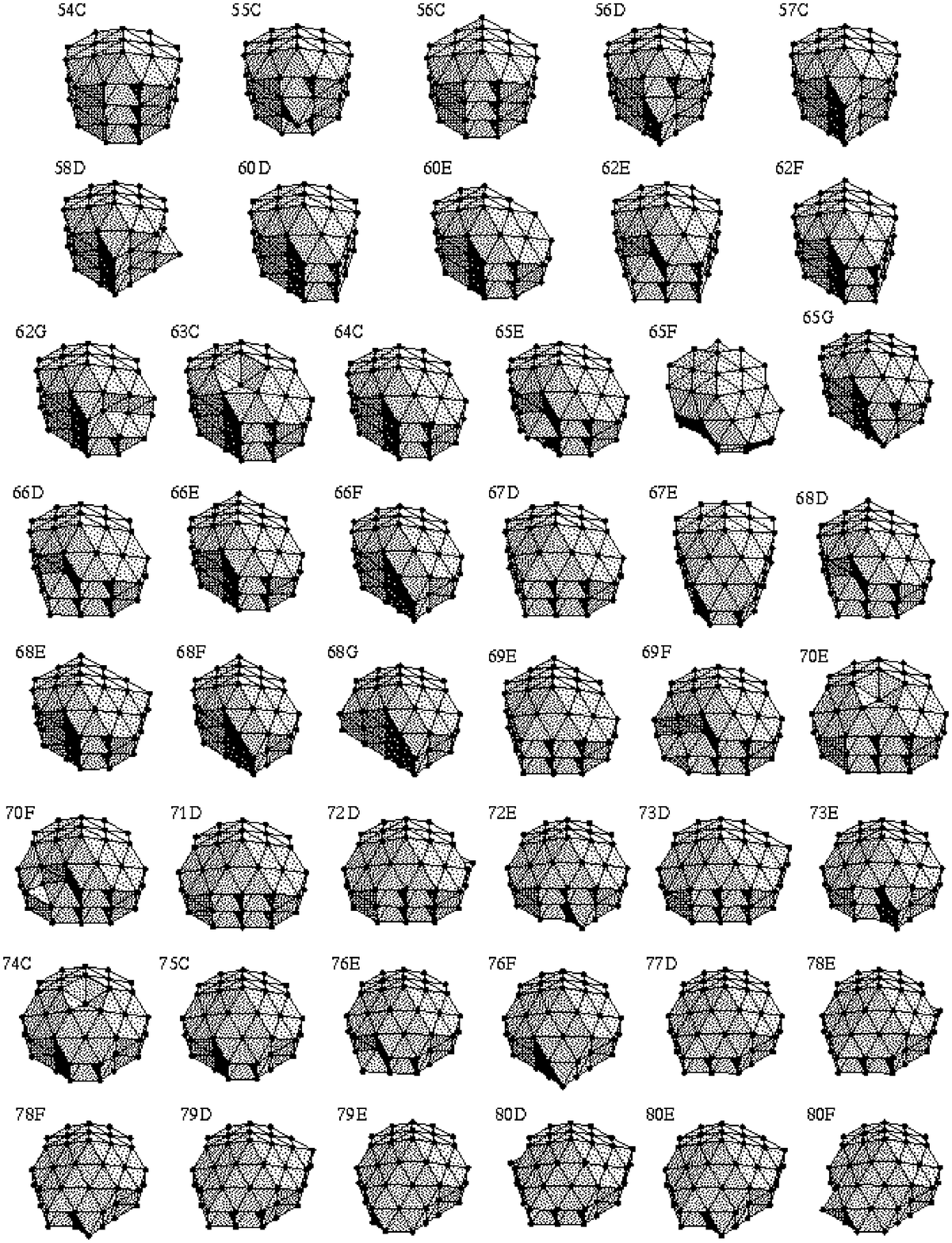,width=17.3cm}
\caption{Global minima based upon a decahedron with five atoms along the 
fivefold axis.}
\label{fig:deca5}
\end{center}
\end{figure}

The decahedral global minima are shown in 
Figs.\ \ref{fig:deca3}--\ref{fig:deca5}.
The structures have been grouped according to the number of atoms along the 
fivefold axis of the pentagonal bipyramid upon which they are based, 
and the decahedral region of the structural phase diagram 
(Fig.\ \ref{fig:phase}) has also been subdivided on this basis.

Decahedral clusters grow by capping exposed $\{100\}$ faces and filling 
in the grooves produced by the re-entrant $\{111\}$ faces.
As this process progresses the structure changes from prolate 
to approximately spherical to oblate.
This cycle begins again when a prolate cluster with a longer decahedral axis 
becomes lower in energy than the oblate cluster (e.g.~at $N\approx 30$ and 54).
For the clusters based upon a pentagonal bipyramid with 5 atoms along the 
fivefold axis ($N\ge 54$), the growth proceeds asymmetrically---the decahedral 
axis does not always pass through the center of the cluster.
For example, for 54C the surface structure of the 75-atom Marks decahedron is 
completed on one side of the cluster before atoms are added to the other.
\end{multicols}
\begin{multicols}{2}

Deviations from this basic growth scheme occur for $N$=21--30 
(Fig.\ \ref{fig:deca3}) and $N$=48--52, 58, 60 and 62 (Fig.\ \ref{fig:deca4}). 
These structures are formed by addition of atoms to the $\{111\}$ faces 
surrounding the fivefold axis in sites which are hcp with respect to the five 
fcc tetrahedra that make up the decahedra.
These structures are more favourable even though they are more strained than 
the usual decahedra because they have a larger $n_{nn}$. 
For $N$=21--30 these structures are actually fragments of the 
55-atom Mackay icosahedron.

The complete Marks decahedron, 75C, is particularly stable.
The value of $\rho_0$ at which it becomes the global minimum (5.81) is 
the lowest of any of the decahedra.
As this value of $\rho_{min}$ suggests, it is also 
the global minimum for \LJ{75}.\cite{JD95c}
This stability is also indicated by the large peak in $\Delta_2E$ for 
$\rho_0$=10 and 14.
Other particularly stable structures occur at $N$=64 and 71; these are 
fragments of 75C with 3 and 4 $\{100\}$ faces of the Marks decahedron complete.

Decahedral structures have been regularly seen in 
supported metal clusters.\cite{Marks94}
However, it is only recently that further experimental evidence for the 
existence of Marks decahedra has been found in studies of gold clusters 
passivated by alkylthiolates.\cite{Alvarez97,Cleveland97a,Cleveland97b}
Whetten and coworkers were able to isolate fractions with narrow size 
distributions which corresponded to the 75-, 101- and 146-atom Marks decahedra.
It is significant to note that our previous paper on Morse clusters\cite{JD95c}
foreshadowed this discovery by recognizing the especial stability of the 
75-atom Marks decahedron, 
thus again showing the utility of Morse clusters as a model system.

\subsection{Close-packed clusters}
\begin{figure}
\begin{center}
\epsfig{figure=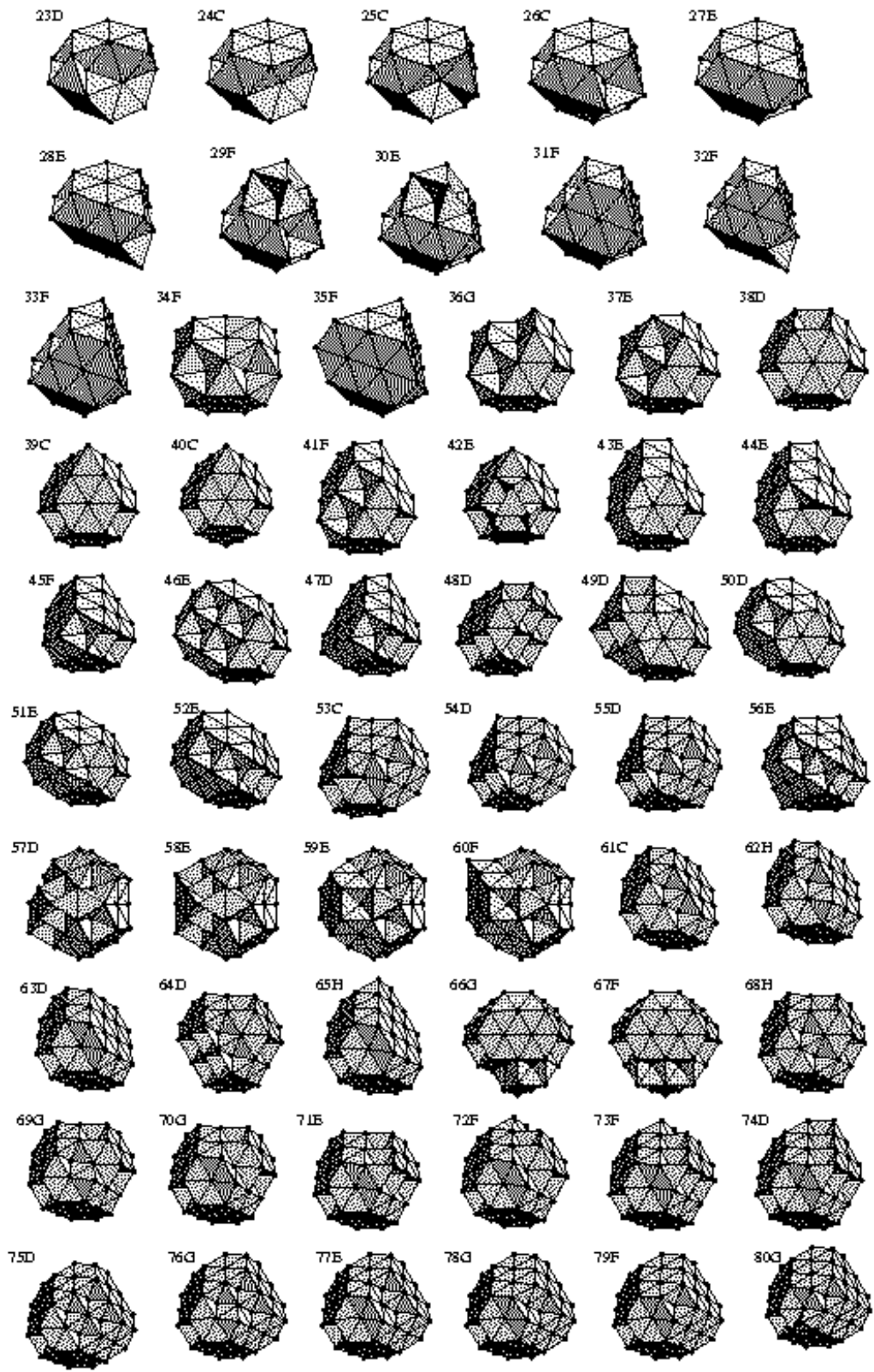,width=14.6cm}
\vglue -0.3 mm
\caption{Global minima based upon close-packing.}
\label{fig:cp}
\end{center}
\end{figure}

The close-packed global minima are 
illustrated in Fig.\ \ref{fig:cp}.
They have a diverse range of structures: there are 4 minima that are fcc, 8 that
are hcp and 46 that involve a mixture of stacking sequences and twin planes.
The preference for close-packed structures with twin planes, even though at many
of the sizes there are fcc isomers with the same number of nearest neighbours, 
occurs for the same reason that bulk hcp has a lower energy than fcc for pair 
potentials, namely a larger energetic contribution from next-nearest neighbours.
The global minima are broadly based on five structures which are especially 
stable: the hcp 26C, the truncated octahedron 38D, the tetrahedral 59E and 
the `twinned truncated octahedra' 50D and 79F.
The latter four give rise to peaks in $\Delta_2 E$ at $\rho_0$=14 
(Fig.\ \ref{fig:D2E}).

The 38-atom truncated octahedron, 38D, is the most stable fcc cluster in the 
size range we consider here. 
It becomes the global minimum at the lowest value of $\rho_0$ (4.76) of any of 
the close-packed structures. 
Curiously, there are two ranges of $\rho_0$ for which it is the global minimum.
At long range, $E_{nnn}$ represents a significant part of the total energy. 
The truncated octahedron is most stable for $4.76<\rho_0<5.40$ because it is 
approximately spherical and so has a larger value of $E_{nnn}$ than the 
more oblate icosahedral structure 38E. 
For shorter-ranged potentials, the contribution of $E_{nnn}$ diminishes 
and so 38E becomes the global minimum for $5.40<\rho_0<6.95$ 
because it has a larger $n_{nn}$. 
Then for $\rho_0>6.95$ the truncated octahedron again becomes the global 
minimum because it has a lower strain energy than 38E. 
There is a growing body of experimental evidence for 
the importance of truncated octahedra.
Parks \etal\ have recently assigned this structure to Ni$_{38}$ by probing 
the cluster's chemical reactivity.\cite{Parks97}
EXAFS (extended x-ray absorption fine structure) spectra of small gold clusters
have been interpreted in terms of the presence of truncated octahedral clusters, 
particularly the 38-atom truncated octahedron.\cite{Pinto}
Gold clusters passivated by alkylthiolate molecules selectively form truncated
octahedra, which can be isolated and formed into 
superlattices.\cite{Whetten96,Andres96}
This structure is also observed for ligated 
38-atom platinum clusters.\cite{Schmid92}

Structures 50D and 79F both have $D_{3h}$ symmetry and a single twin plane 
passing through the structure. 
The two halves of the structure have the surface morphology of truncated octahedra. 
Indeed 79F can be formed from the 79-atom truncated octahedron by rotation of 
one half of the structure
by 60$^\circ$ about an axis perpendicular to one of the $\{111\}$ planes. 
Both structures have the same number of nearest neighbours and 79F is slightly 
lower in energy only because of the larger contribution from next-nearest 
neighbours that results from the twin plane.
Again there is recent experimental evidence for the stability of this type of 
structure for gold clusters,\cite{Cleveland97a} 
although at larger sizes ($N$=225 and 459) 
than considered in this study.

The closed-packed structures from $N$=57--60 are based on 59E.
This structure is a 31-atom truncated tetrahedron with the faces covered by 
four seven-atom hexagonal overlayers occupying hcp sites with respect to 
the underlying tetrahedron.
The stability of 59E comes from the combination of its high proportion of 
$\{111\}$ faces and its spherical shape. 
If atoms are added to one of the grooves in 59E a decahedral-like
axis results, indicating a possible path between decahedral and 
closed-packed clusters.
\end{multicols}

\begin{multicols}{2}

It is worth noting that the cuboctahedron is not the lowest energy close-packed 
structure for $N$=55.
In fact, it has four fewer nearest neighbours than structure 55D. 
Hence, when magic numbers occur at sizes corresponding to both complete 
Mackay icosahedra and cuboctahedra\cite{Martin90,Martin91a} 
($N$=13, 55, 147, ...) it is more likely that they are due to icosahedra.
Furthermore, one has to interpret with caution those studies which seek 
to find the relative stability of fcc and icosahedral structures by comparing 
cuboctahedra with Mackay icosahedra\cite{Xie,Up92,Wang,Lim}
because the cuboctahedra are likely to be suboptimal fcc structures. 

Although this conclusion does not simply carry over to clusters surrounded by 
ligand shells---the ligands could significantly modify the relative surface 
energies of $\{111\}$ and $\{100\}$ faces---it is interesting to note that a 
recent reinvestigation of clusters which were originally thought to be 
cuboctahedral 55-atom gold clusters\cite{Schmid81,Wallenberg85}
seems to disprove this structural assignment.\cite{Rapoport97}

\subsection{Structures corresponding to long range}
\begin{figure}
\vglue -4mm
\begin{center}
\epsfig{figure=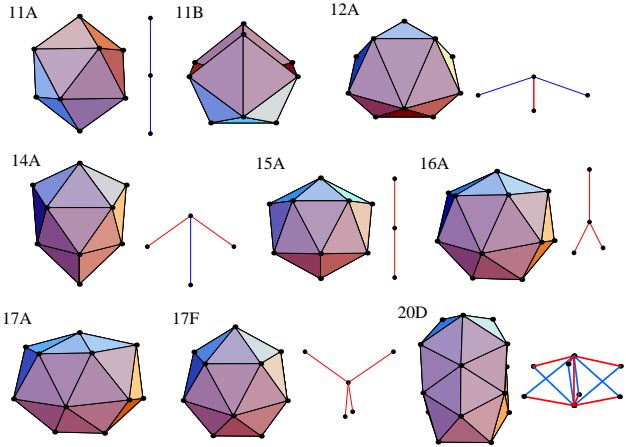,width=8.6cm}
\vglue -1mm
\begin{minipage}{8.5cm}
\caption{Low energy minima for small clusters at low values of $\rho_0$. 
If appropriate, the disclination network is displayed next to each structure.
$-72^\circ$ disclinations are represented by red lines and 
$+72^\circ$ disclinations by blue lines.
17F and 20C are never global minima.}
\end{minipage}
\label{fig:dis.small}
\end{center}
\end{figure}

In this subsection, we consider those 
structures which become the global minimum only at low values of $\rho_0$.
We have restricted our study to those clusters with $\rho_0\ge 3$, 
since we do not know of a case where longer-ranged potentials might be relevant.
The low $\rho_0$ structures that we have found separate into two size ranges: 
those with $N$ around 13 and those with $N>45$. 
The former set have been described in a previous paper,\cite{JD95c}
but we illustrate them again in Fig. \ref{fig:dis.small} because the connection
to Kasper polyhedra\cite{FrankK58,FrankK59} was not originally identified and 
because they are important for understanding the structures that occur at larger $N$.

The majority of the structures associated with low $\rho_0$ are polytetrahedral: 
the entire cluster can be divided into tetrahedra with atoms at the vertices. 
These minima also tend to be close to spherical in shape and highly strained in order to maximize $n_{nn}$. 
As we noted earlier, the 13-atom icosahedron is polytetrahedral. 
In this case, each nearest-neighbour contact between the centre and a vertex is 
the common edge of five tetrahedra. This is also true of all the nearest-neighbour contacts in
the rhombic tricontahedron (45A) which do not lie on the surface.
Nearest-neighbour contacts which are surrounded by more or less than five tetrahedra are said 
to have defects called {\it disclination lines\/} running along the interatom vector.
Those contacts surrounded by more than five tetrahedra are termed negative disclinations (if there are six
it is a $-72^\circ$ disclination, if there are seven a $-144^\circ$ disclination, \dots)
and those surrounded by fewer than five tetrahedra are termed positive disclinations (if there are four
it is a $+72^\circ$ disclination and if there are three a $+144^\circ$ disclination).

Most of the structures associated with low $\rho_0$ are 
polytetrahedral and involve disclinations. 
Although packing five tetrahedra around a nearest-neighbour 
contact involves some strain, the energetic penalty associated with either 
more or less tetrahedra is greater.
Therefore, structures involving disclinations are only likely to be lowest 
in energy for long-ranged potentials where the associated strain can be 
most easily accommodated, and where they must have
a larger $n_{nn}$ than the alternative disclination-free structures.
A $-72^\circ$ disclination line involves less strain than a 
$+72^\circ$ disclination line because of the gap that remains when five regular 
tetrahedra share a common edge (Fig.\ \ref{fig:gaps}(a)).
Consequently, structures which involve only negative disclinations, 
or an excess of them, are more common amongst the low $\rho_0$ global minima.

To visualize the network of disclination lines in a structure, 
one must first partition space according to the Voronoi procedure, in which 
each point is assigned to the Voronoi polyhedron of the atom to which it is closest.
This allows nearest neighbours to be defined as those atoms whose 
Voronoi polyhedra share a face.
The Delaunay network that results from joining all such nearest neighbours
is the dual of the Voronoi construction and divides all space into tetrahedra.
This definition of a nearest neighbour has been termed geometric,
rather than physical (e.g.~using a cutoff distance), and the corresponding 
division of space into tetrahedra is artificial in the sense that it is 
independent of whether a polytetrahedral description is appropriate.

In practice we determined the Voronoi polyhedra from the fact that a set of four atoms
constitutes a Delaunay tetrahedron if the sphere touching all four atoms contains no other atoms.\cite{Ashby}
The centre of this sphere is then a vertex of the Voronoi polyhedron of each atom.
As the number of tetrahedra around a nearest-neighbour contact is the same as the number
of sides for the face common to the Voronoi polyhedra of both nearest neighbours, 
disclination lines can then be assigned. 
However, problems can occur in assigning the Delaunay network if there are
more than four atoms exactly on the surface of the sphere.
Such a degeneracy, which only occurs as a result of symmetry, renders the analysis non-unique. 
This is the case for the 55-atom Mackay icosahedron and for bulk close-packed solids
because of the presence of octahedral interstices, but it is not a problem here 
since we only apply the method to clusters that are polytetrahedral in character.
One further consideration is that the analysis should not be applied to nearest-neighbour contacts
between the surface atoms of a cluster.

The smallest global minimum that involves a disclination line is 11A
where the central atom is surrounded by a 10-atom coordination shell. 
This encapsulation gives the structure a larger $n_{nn}$ 
than the incomplete icosahedron (structure 11B of ref.\ \onlinecite{JD95c}), 
but results in a larger strain energy.\cite{JD95c}
The structure involves a single positive disclination line running through the centre of the cluster.
Similarly, for $N$=12 and 14--16, clusters with a single coordination shell 
become lower in energy than structures based on the icosahedron at long range.
Structures 11A, 12A, 14A, 15A, 16A and 17F (the second lowest energy structure of \M{17} at $\rho_0=3$)
correspond to Kasper polyhedra. 
Of these structures 15A is the most stable in terms of the range of $\rho_0$ for which it is 
the global minimum.\cite{JD95c} 
The Kasper polyhedra are the deltahedral coordination shells that involve the 
minimum number of disclinations. 
They are important in the Frank-Kasper phases,\cite{FrankK58,FrankK59}
which are crystalline structures that are polytetrahedral 
and involve ordered arrays of negative disclination lines.
Much interest has been focussed on the Frank-Kasper phases because they are closely related to 
icosahedral quasicrystals.\cite{Henley}
Indeed a recent three-dimensional model of quasicrystalline structure 
was based upon clusters involving disclination lines similar to those we find here.\cite{Borodin}

\begin{figure}
\begin{center}
\epsfig{figure=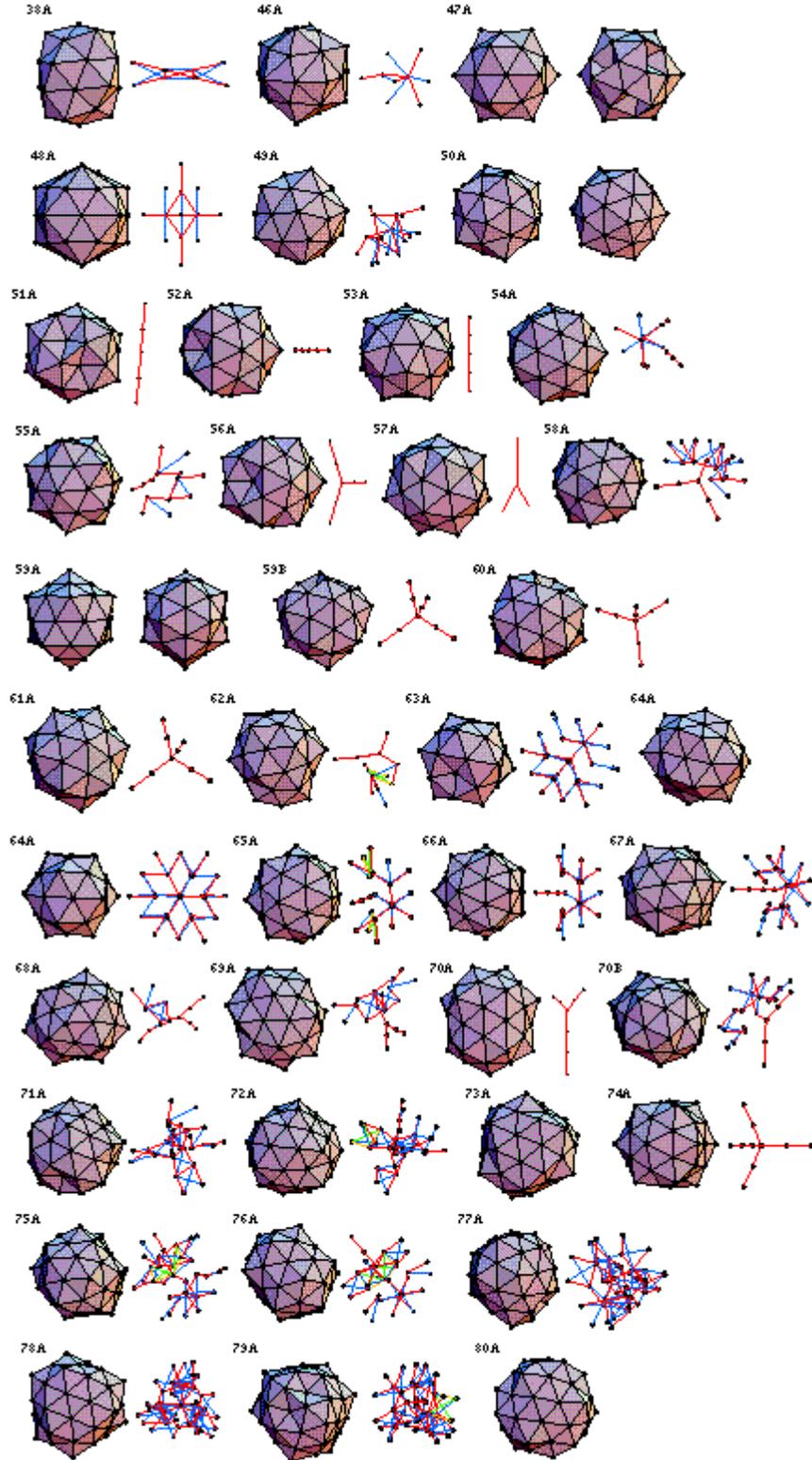,width=12.3cm}
\vglue -0.7mm
\caption{Global minima at low values of $\rho_0$ for $N\ge 38$.
If appropriate, the disclination network is displayed next to each structure.
$-72^\circ$ disclinations are represented by red lines, 
$+72^\circ$ disclinations by blue lines,
$-144^\circ$ disclinations by green lines and 
$+144^\circ$ disclinations by yellow lines.}
\label{fig:dis.large}
\end{center}
\end{figure}

Structures 11A, 12A, 14A, 15A and 16A are all deltahedral, and so growth
can occur in both `anti-Mackay' and `Mackay' sites (Fig.\ \ref{fig:over}). 
However, if one considers the addition of a hexagonal pyramidal cap to the most stable of the
Kasper polyhedra, 15A, the result is a $D_{6h}$ structure with a positive disclination running the length
of the symmetry axis (a disclinated equivalent of the double icosahedron 19A) which has only the same
$n_{nn}$ as the icosahedral structure 22A. 
Disclinated polytetrahedra cannot compete with the disclination-free polytetrahedral 
structures produced by an anti-Mackay overlayer on the icosahedron. 
Only once the latter growth sequence is completed at $N$=45 are structures with disclinations
again global minima (Fig.\ \ref{fig:dis.large}).
The one exception is structure 38A, which is similar to the icosahedral structures 38B and 38C,
but has two positive disclinations running through the structure in a strange double helical twist.

Interestingly, structures with anti-Mackay growth on 11A have been recently 
observed for M$^+$RG$_N$ (M---metal, RG---rare gas) clusters where the metal ion is
sufficiently small with respect to the rare gas atoms.\cite{Luder97} 
The metal ion presumably lies in the centre of the cluster, and the size ratio 
ensures that the cluster is least strained when the metal ion is surrounded by 10 rare
gas atoms.
Similarly, one might expect structures with negative disclinations to occur for
AB$_N$ clusters when A is sufficiently large with respect to B that a coordination number of 
larger than twelve is favoured for the A atom.

The most stable disclinated polytetrahedral structures occur at $N$=53, 57, and 61 (Fig.\ \ref{fig:D2E}(a)). 
These are the sizes for which complete `anti-Mackay' overlayers on 15A, 16A and 17F are possible.
The effect of the overlayers is to extend the disclination lines emanating from the central atom.
In these structures, those interior atoms not lying on a disclination line are icosahedrally coordinated.
Many other minima are related to these stable structures: 
51A and 52A are based on 53A but with missing vertex atoms; 
similarly, 59B and 60A are based on 61A; and 70A and 74A are based on 57A and 61A, respectively, 
but with an additional 13-atom cap which extends one of the `arms' of the disclination network.
Furthermore, many of the other structures include parts of the disclination networks of 53A, 57A and 61A, 
but combined with a more disordered array of disclinations in another part of the cluster.

Other interesting structures are also seen. 
47A, 50A and 59A seem to have a mixture of Mackay and anti-Mackay overlayers.
64A is formed from the rhombic tricontahedron 45A by the extension of the structure along a three-fold axis
and the addition of a ring of atoms in the centre.
In the middle of 64A is the $D_{3d}$ structure 20D (the third lowest energy structure for \M{20} at $\rho_0$=3)
which can be regarded as two highly strained interpenetrating icosahedra.
However, for the larger clusters it becomes difficult to recognize any structural motifs,
and some just seem to be disordered tangles of disclinations. 

The above results are particularly interesting because of their relevance to 
our understanding of liquid structure. 
The minima described above, when reoptimized at larger values of $\rho_0$, 
correlate with structures which lie in the lower energy range of the band of 
minima associated with liquid-like clusters.\cite{JD96a,JD96b}
This is because simple liquids have significant polytetrahedral 
character,\cite{NelsonS} as has been shown by the success of the dense random 
packing of hard spheres\cite{Bernal60,Bernal64} as a model for metallic 
glasses\cite{Cargill} and later by computer simulations.\cite{Jonsson}
Indeed, Nelson has suggested that simple liquids are polytetrahedral packings
that are characterized by a disordered arrangement of disclination 
lines.\cite{Nelson83a,Nelson83b}
Consequently, by examining the global minima associated with low
$\rho_0$ we can study the size evolution of polytetrahedral packings,
and the development of bulk liquid structure.

\end{multicols}
\begin{multicols}{2}
At small sizes disclination-free polytetrahedra based on icosahedra with 
anti-Mackay overlayers are possible; at larger sizes polytetrahedra with 
ordered arrangements of disclinations are most common, and finally at the largest
sizes in this study the polytetrahedra have a disordered disclination network. 
The latter geometries are structurally very similar to fragments of bulk liquid, 
except that the density of disclination lines is lower.\cite{JD96b}

Theoretical studies of sodium clusters have shown that
amorphous structures are lower in energy than regular geometries up to at 
least 340 atoms, the largest size considered in that study.\cite{Glossman}
The present results suggest that the these disordered structures are due 
to the relatively long range of the sodium interatomic potential.
Amorphous structure for sodium is also suggested by the experimental 
observation of electronic shells\cite{Knight} and 
supershells,\cite{Pedersen} which are incompatible with any 
of the ordered morphologies one might expect.\cite{Pavloff}
Indeed, the transition from electronic to geometric magic numbers which occurs
at about 1000 atoms\cite{Martin90} probably reflects a change in the lowest 
energy structures from amorphous minima to Mackay icosahedra, which can be 
explained by the twin effects of the size and the range of the 
potential.\cite{JD96a,JD96b}

Since most minima associated with small values of $\rho_0$ do not have 
a common lattice or packing scheme, 
it is difficult to predict the low energy structures, and so most have been found by 
one of the global optimization methods.
Fortunately, these methods are most likely to succeed for long range potentials 
because the corresponding PES's are smoother and support fewer minima 
than for shorter range.
Another consequence of the lack of lattice structure is that the differentiation 
between nearest neighbours and next-nearest neighbours becomes ambiguous and 
the decomposition of eqn.\ (2) more arbitrary. For these clusters we chose to 
define $n_{nn}$ using a cutoff at $\rho_0$=4 which corresponds to the distance 
at which the pair energy is $0.6\,\epsilon$. 

\section{Discussion}In this paper we have attempted to find the global 
minima for Morse clusters as a function of $\rho_0$ and the number of atoms. 
The global potential energy minimum represents the equilibrium structure 
at zero Kelvin, but to predict the structure at non-zero temperatures 
we must consider the free energy, and the effect of other low energy minima.
This can be done by summing the density of states over all 
the relevant minima.\cite{JD95a}
An illustration of this approach for understanding the structure of Morse 
clusters has been given previously,\cite{JD95c,JD95cnote}
revealing that for \M{75} at $\rho_0$=6, the equilibrium structure changed 
from decahedral to icosahedral at very low temperature. 
This transition is simply a consequence of the larger 
entropy for the icosahedral region of configuration space---there are far more
low energy icosahedral minima.

This effect is likely to be general. 
At the magic numbers for a morphology the entropy will be low because there is a single 
unique low entropy structure and a large gap to other isomers with the same morphology. 
Therefore the finite temperature equivalent of the structural phase diagram of Fig.\ \ref{fig:phase}
is likely to show weakened magic number effects and so have smoother boundaries between the 
different morphologies.
Furthermore, the energy gap between the lowest energy ordered minimum and the liquid-like band of minima
increases as the range becomes shorter, and hence the melting temperature increases with $\rho_0$.\cite{JD96b}
Therefore the region of the phase diagram where disordered polytetrahedral structures have the lowest free 
energy is likely to spread up from the bottom as the temperature increases.

In this paper we have considered only isotropic pairwise additive interactions.
As noble gas clusters and clusters of \csixty\ molecules can be reasonably 
modelled by such potentials we would expect the structures we have found at the appropriate 
values of $\rho_0$ to be similar to the actual structures of these clusters. 
Our results lead us to predict that neutral clusters of \csixty\ molecules exhibit
decahedral and fcc structures at small sizes because of the short range of the intermolecular potential.
This basic conclusion has been confirmed in studies using more realistic potentials.\cite{JD96d,JD97c}

In contrast, making predictions for metal clusters is problematic
because the range of the potential is only one factor determining the structure 
and many-body terms, in particular, may also be important.\cite{Wales90a,JD92}
These terms may affect the relative surface energies of $\{111\}$ and $\{100\}$ faces, 
and so alter the energetic competition between icosahedral, decahedral and fcc structures.\cite{Up92}
For example, in a study of lead clusters cuboctahedra are always found to be lower in energy 
than icosahedra because the surface energies of $\{111\}$ and $\{100\}$ faces are nearly equal.\cite{Lim}

Nevertheless our results are of value to the field of metal cluster structure. 
Firstly, they enable particularly stable structural forms to be identified. 
For example, in our previous paper on Morse clusters we identified the 38-atom
octahedron and the 75-atom Marks decahedron as particularly stable.
Subsequently, they have both been observed experimentally;\cite{Parks97,Alvarez97}
it even being possible to isolate fractions of the latter when passivated by surfactants.
This correspondence between the Morse structures and those of real systems encourages
us to believe that some of the general principles that determine stability in our
simple model system do carry over to real clusters.

Secondly, the Morse structural database should be useful in providing 
candidate structures for comparison with the indirect structural information 
yielded by experiments on size-selected clusters. 
Finally, the database can also provide plausible starting structures 
for theoretical studies with more realistic, 
but computationally expensive, descriptions of the interactions; 
this expense would prevent
the type of extensive searches that have been performed in this paper.
Indeed, we have used the database in this way in studies of 
metal clusters modelled by the Sutton-Chen family of potentials\cite{WalesD97b}
and clusters of \csixty\ molecules.\cite{JD96d,JD97c}
For these reasons, the coordinates for all the global minima given in this 
and previous papers\cite{JD95c,WalesD96} will be made available 
on the world-wide-web.\cite{Web}

\section{Conclusion}We have shown how the range of an isotropic pairwise 
additive potential determines the structure of atomic clusters.
In particular, we have identified four principal structural regimes.
For long-ranged potentials at sizes $N=10$--18 and $N>45$ the global minima are 
generally polytetrahedral and can be analyzed in terms of disclination lines.
For the smaller clusters the disclinations form ordered arrangements and 
the structures are fragments of bulk Frank-Kasper phases. 
As the size increases it becomes more likely that the disclination network 
is a disordered tangle; these global minima have amorphous structures similar 
to those exhibited by liquid-like clusters. 
At intermediate ranges of the potential icosahedra are dominant. 
As the range decreases, first decahedral and 
then fcc structures become lowest in energy. 
These trends have been explained by considering the strain energies and 
the number of nearest neighbour contacts associated with each regime. 
The effect of decreasing the range of the potential is to destabilize 
the strained structures. 

\acknowledgements
We are grateful to the Engineering and Physical Sciences Research Council 
(J.P.K.D.) and the Royal Society (D.J.W.) for financial support.
\end{multicols}
\begin{table}
\caption{Global minima of M$_N$ for $N\le 80$. 
Energies at values of $\rho_0$ for which the structure is 
the global minimum are given in bold. 
$\rho_{min}$ and $\rho_{max}$ give the range of $\rho_o$ for 
which the structure is the global minimum. 
If at a particular value of $\rho_0$ a structure is not a minimum 
but a higher order saddle point, the index of the stationary point 
(the number of negative eigenvalues of the
Hessian) is given in square brackets after the energy.
$E_{\rm strain}$ has been calculated at $\rho_0$=10. 
If a structure is not stable at $\rho_0$=10 no value of 
$E_{\rm strain}$ is given. All energies are given in $\epsilon$.   
}
\begin{center}
\begin{tabular}{cccrccccrr}
 &  PG & $n_{nn}$ & $E_{\rm strain}$ & $\rho_0=3$ & $\rho_0=6$ & $\rho_0=10$ \ & 
$\rho_0=14$ \ & $\rho_{min}$ & $\rho_{max}$ \\
\hline
 5A & $D_{3h}$ & 9 & 0.000 & {\bf\,~-9.299500} & {\bf\,~-9.044930} & {\bf~\,-9.003565} & {\bf\,~-9.000283} & & \\
 6A & $O_h$ & 12 & 0.000 & {\bf -13.544229} & {\bf -12.487810} & {\bf -12.094943} & {\bf -12.018170} & & \\
 7A & $D_{5h}$ & 16 & 0.062 & {\bf -17.552961} & {\bf -16.207580} & {\bf -15.956512} & {\bf -15.883113} & & \\
 8A & $D_{2d}$ & 18 & 0.006 & {\bf -22.042901} & -19.161862 & -18.275118 & -18.076248 & & 5.28 \\
 8B & $C_s$ & 19 & 0.062 & & {\bf -19.327420} & {\bf -18.964638} & {\bf -18.883688} & 5.28 & \\
 9A & $D_{3h}$ & 21 & 0.002 & {\bf -26.778449} & -22.330837 & -21.213531 & -21.037957 & & 3.42 \\
 9B & $C_{2v}$ & 23 & 0.186 & -26.607698 & {\bf -23.417190} & {\bf -22.850758} & {\bf -22.644892} & 3.42 & \\
 10A & $D_{4d}$ & 24 & 0.002 & -31.519768 & -25.503904 & -24.204958 & -24.031994 & & 2.28 \\
 10B & $C_{3v}$ & 27 & 0.694 & {\bf -31.888630} & {\bf -27.473283} & {\bf -26.583857} & {\bf -26.132735} & 2.28 & \\
 11A & $D_{4d}$ & 34 & 10.374 & {\bf -37.930817} & -28.795153[4] & -23.666072[5] & & & 3.40 \\
 11B & $C_{2v}$ & 32 & & -37.891674[1] & & & & 3.40 & 3.67 \\
 11C & $C_{2v}$ & 31 & 0.792 &  & {\bf -31.521880} & {\bf -30.265230} & -29.588130[1] & 3.67 & 13.57 \\
 11D & $C_s$ & 30 & & & & & {\bf -29.596054} & 13.57  & 15.29 \\
 11E & $C_2$ & 30 & 0.248 & -36.613190[1] & -30.698890 & -29.808994 & -29.524398 & 15.29 & 20.60 \\
 11F & $C_{2v}$ & 29 & 0.001 & -36.697760 & -30.431713 & -29.215510 & -29.037941 & 20.60 & \\
 12A & $C_{2v}$ & 38 & & -43.971339[1] & -35.199881[1] & & & & 2.63 \\
 12B & $C_{5v}$ & 36 & 1.704 & {\bf -44.097880} & {\bf -36.400278} & {\bf -34.366755} & -33.115942[1]  & 2.63 & 12.15 \\
 12C & $C_s$ & 34 & & & & & -33.199505 & 12.15 & 13.03 \\
 12D & $D_{2d}$ & 34 & 0.346 & -41.816393 & -34.838761 & -33.724155 & {\bf -33.332305} & 13.03 & 17.08 \\
 12E & $D_{3h}$ & 33 & 0.001 & -42.121440 & -34.568002 & -33.222331 & -33.038298 & 17.08 & \\
 13A & $I_h$ & 42 & 2.425 & {\bf -51.737046} & {\bf -42.439863} & {\bf -39.662975} & {\bf -37.258877} &  & 14.76 \\
 13B & $D_{5h}$ & 37 & 0.141 &  -49.998058[1] & -39.360710[1] & -37.208019[1] & -36.790507 & 14.76 &\\
 14A & $C_{2v}$ & 46 & 4.373 & {\bf -56.754744} & -44.827522[1] & -41.717043[1] & & & 3.23 \\
 14B & $C_{3v}$ & 45 & 2.425 &  -56.660471 & {\bf -45.619277} & {\bf -42.675222} & -40.259823 & 3.23 & 13.06 \\ 
 14C & $C_{2v}$ & 41 & 0.141 & -55.971620[2] & -43.634048[1] & -41.249282 & {\bf -40.798348} & 13.06 & \\
 15A & $D_{6d}$ & 50 & 9.527 & {\bf -63.162119} & -47.570579 & -40.569211[10] & -35.758904[13] & & 3.72 \\
 15B & $C_{2v}$ & 49 & 2.573 & -62.593904 & {\bf -49.748409} & {\bf -46.541404} & -44.086633 & 3.72 & 12.71 \\
 15C & $C_{2v}$ & 45 & 0.141 & -62.631372[2] & -47.952559 & -45.293844 & {\bf -44.806437} & 12.71 & \\
 16A & $D_{3h}$ & 54 & 11.222 & {\bf -69.140648} & -50.834213 & -42.887569[12] & & & 3.39 \\
 16B & $C_s$ & 53 & 2.868 & -68.757203 & {\bf -53.845835} & {\bf -50.261947} & -47.831957 & 3.39 & 11.99 \\
 16C & $C_{2v}$ & 49 & 0.142 & -68.575718[1] & -52.265348 & -49.338173 & {\bf -48.814517} & 11.99 & \\
\end{tabular}
\end{center}
\end{table}
\addtocounter{table}{-1}
\begin{table}
\caption{continued.}
\vglue 2mm
\begin{tabular}{cccrccccrr}
 &  PG & $n_{nn}$ & $E_{\rm strain}$ & $\rho_0=3$ & $\rho_0=6$ & $\rho_0=10$ \ & 
$\rho_0=14$ \ & $\rho_{min}$ & $\rho_{max}$ \\
\hline
 17A & $D_{3h}$ & 58 & & {\bf -75.662417} & -53.156042[2] & & & & 3.42 \\
 17B & $C_{3v}$ & 57 & 3.372 & -75.147372 & -57.884517 & -53.772213 & -51.329560 & 3.42 & 4.88 \\
 17C & $C_s$ & 57 & 3.281 & -75.091367 & -57.912963 & -53.862044 & -51.440588 & 4.88 & 4.91 \\
 17D & $C_2$ & 57 & 3.163 & -75.005403 & {\bf -57.941386} & {\bf -53.983559} & & 4.91 & 11.30 \\
 17E & $C_{2v}$ & 53 & 0.142 & -74.868921[1] & -56.573571 & -53.382277 & {\bf -52.822588} & 11.30 & \\
 18A & $C_2$ & 65 & & {\bf -82.579266} & -59.881449[1] & & & 2.14 & 3.03 \\
 18B & $C_{5v}$ & 62 & 4.500 & -82.548885 & {\bf -62.689245} & {\bf -57.657135} & -54.059707 & 3.03 & 10.13 \\
 18C & $C_s$ & 61 & 3.528 & -81.256639 & -62.002920 & -57.634324 & -55.776126 & 10.13 & 10.43 \\
 18D & $D_{5h}$ & 57 & 0.142 & -81.490185 & -60.926500 & -57.429683 & {\bf -56.830907} & 10.43 & \\
 19A & $D_{5h}$ & 68 & 6.001 & {\bf -90.647461} & {\bf -68.492285} & {\bf -62.166843} & -56.676685[4] & & 10.70 \\
 19B & $C_{2v}$ & 61 & 0.151 & -87.485744 & -65.064771 & -61.427105 & {\bf -60.812425} & 10.70 & \\
 20A & $C_{2v}$ & 72 & 6.507 & {\bf\ \,-97.417393} & {\bf -72.507782} & {\bf -65.679115} & -61.327229 & 2.02  & 10.24 \\
 20B & $C_{2v}$ & 65 & 0.162 & \ \,-94.222416 & -69.202704 & -65.423697 & {\bf -64.791953} & 10.24 & \\
 21A & $C_{2v}$ & 76 & 7.133 & {\bf -104.336946} & -76.487266 & -69.068687 & -65.179591[1] & & 5.40 \\
 21B & $C_1$ & 76 & 6.922 & -104.004129 & {\bf -76.529139} & -69.276346 & -65.778898 & 5.40 & 9.84 \\
 21C & $C_s$ & 69 & 0.169 & & -73.577014 & {\bf -69.449904} & {\bf -68.783571} & 9.84 & \\
 22A & $C_s$ & 81 & 8.198 & {\bf -112.041223} & {\bf -81.136735} & -73.014321 & -68.580862[1] &  & 9.58 \\
 22B & $C_1$ & 73 & 0.169 & & -77.887855 & {\bf -73.494292} & {\bf -72.791747} & 9.58 & \\
 23A & $D_{3h}$ & 87 & 10.597 & {\bf -120.786879} & {\bf -86.735494} & -76.630624 & -70.816059[5] & & 8.35 \\
 23B & $D_{3h}$ & 84 & 6.097 & -116.279438 & -84.940552 & -78.143867 & -74.442341[2] & 8.35 & 9.71 \\
 23C & $C_s$ & 78 & 0.433 & & -83.504908 & {\bf -78.325380} & {\bf -77.302495} & 9.71 & 22.45 \\
 23D & $C_1$ & 76 & 0.007 & & -82.252747 & -76.898804 & -76.157457 & 22.45 & \\
 24A & $C_s$ & 91 & 10.947 & {\bf -127.884549} & {\bf -90.685398} & -80.295459[1] & -74.613738[2] & 2.79  & 8.55 \\
 24B & $C_1$ & 82 & 0.434 & & -87.820376 & {\bf -82.370214} & {\bf -81.309508} & 8.55 & 15.34 \\
 24C & $C_{2v}$ & 81 & 0.007 & & -87.626843 & -81.941420 & -81.164168 & 15.34 & \\
 25A & $C_s$ & 96 & 12.090 & {\bf -136.072704} & {\bf -95.127899} & -84.168765 & & 2.62 & 7.76 \\
 25B & $C_{3v}$ & 87 & 0.893 & & -93.342771 & {\bf -86.989688} & {\bf -85.477376} & 7.76 & 15.40 \\
 25C & $C_s$ & 85 & 0.010 &  & -92.241466 & -86.015371 & -85.176789 & 15.40 & \\
 26A & $T_d$ & 102 & 15.393 & {\bf -145.322134} & {\bf -100.549598} & -86.882333[11] & & & 7.89 \\
 26B & $C_{2v}$ & 91 & 0.491 & -138.940920[3] & -97.363225 & {\bf -91.370250} & {\bf -90.210764} & 7.89 & 14.20 \\
 26C & $D_{3h}$ & 90 & 0.008 & & -97.648652 & -91.085136 & -90.189274 & 14.20 & \\
 27A & $C_s$ & 106 & & {\bf -152.513867} & -104.489430 & & & & 4.72 \\
 27B & $C_{2v}$ & 106 & & -151.734925 & {\bf -104.745275} & & & 4.72 & 7.71 \\
 27C & $C_s$ & 97 & 3.076 & & -102.749592 & -94.736230 & -91.236563[1] & 7.71 & 8.53 \\
 27D & $C_s$ & 95 & 0.491 & -146.171281[2] & -101.722918 & {\bf -95.419490} & {\bf -94.219798} & 8.53 & 14.22 \\
 27E & $C_s$ & 94 & 0.008 & & -101.920210 & -95.121297 & -94.195553 & 14.22 & \\
 28A & $C_s$ & 111 & & {\bf -160.773356} & -108.854564 & & & & 4.82 \\
 28B & $C_s$ & 111 & & -160.385239 & {\bf -108.997831} & & & 4.82 & 6.70 \\
 28C & $C_{3v}$ & 102 & 3.369 & & -108.186446 & -99.524026 & -95.692056 & 6.70 & 8.88 \\
 28D & $C_s$ & 100 & 0.983 & & -107.213896 & {\bf -100.008358} & {\bf -98.331711} & 8.88 & 14.57 \\
 28E & $C_{2v}$ & 98 & 0.008 & & -106.238844 & -99.161073 & -98.202117 & 14.57 & \\
 29A & $D_{3h}$ & 117 & & {\bf -170.115560} & {\bf -114.145949} & & & & 6.90 \\
 29B & $C_1$ & 106 & 3.370 & & -112.655980 & -103.586316 & -99.702681 & 6.90 & 8.95 \\
 29C & $C_1$ & 104 & 0.988 & & -111.543685 & {\bf -104.051412} & -102.333194 & 8.95 & 11.07 \\
 29D & $C_{2v}$ & 103 & 0.246 & & -111.508961 & -103.946446 & -102.743899 & 11.07 & 11.17 \\
 29E & $D_{5h}$ & 103 & 0.225 & -164.720567 & -111.353973 & -103.926827 & {\bf -102.774589} & 11.17 & 21.90 \\
 29F & $C_{3v}$ & 102 & 0.010 & & -111.135014[1] & -103.298583 & -102.227015 & 21.90 \\
 30A & $C_{2v}$ & 121 & & {\bf -177.578647} & -118.115802 & & & & 4.47 \\
 30B & $C_{2v}$ & 121 & 8.279 & -176.940971 & {\bf -118.432844} & -103.407917[1] & & 4.47 & 6.86 \\
 30C & $C_{2v}$ & 109 & 1.538 &  & -117.010672 & {\bf -108.571179} & -106.426033 & 6.86 & 12.61 \\
 30D & $C_s$ & 107 & 0.238 & & -115.940968 & -107.985497 & {\bf -106.765372} & 12.61 & 21.33 \\
 30E & $C_s$ & 106 & 0.012 & & -115.625207 & -107.367696 & -106.239394 & 21.33 & \\ 
\end{tabular}
\end{table}
\addtocounter{table}{-1}
\begin{table}
\caption{continued.}
\vglue 2mm
\begin{tabular}{cccrccccrr}
 &  PG & $n_{nn}$ & $E_{\rm strain}$ & $\rho_0=3$ & $\rho_0=6$ & $\rho_0=10$ \ & 
$\rho_0=14$ \ & $\rho_{min}$ & $\rho_{max}$ \\
\hline
 31A & $C_1$ & 126 & & {\bf -185.984248} & -122.440052 & & & & 4.60 \\
 31B & $C_s$ & 126 & & -185.299446 & {\bf -122.857743} & & & 4.60 & 6.35 \\
 31C & $C_s$ & 115 & 3.490 & & -122.342421 & -112.542723 & -108.722594 & 6.35 & 8.93 \\
 31D & $C_{2v}$ & 112 & 0.269 & -184.500261 & -121.523693 & {\bf -113.066115} & -111.724317 & 8.93 & 10.98 \\  
 31E & $C_{2v}$ & 112 & 0.246 & -181.901112[1] & -121.367441 & -113.048378 & {\bf -111.760670} & 10.98 & 20.90 \\
 31F & $C_s$ & 111 & 0.011 & & -120.805231 & -112.376045 & -111.239641 & 20.90 & \\
 32A & $C_{2v}$ & 132 & & {\bf -195.468461} & -127.643751 & & & & 5.93 \\
 32B & $C_{2v}$ & 120 & 3.831 & & {\bf -127.771395} & {\bf -117.284334} & & 5.93 & 10.27 \\
 32C & $C_s$ & 116 & 0.269 & & -125.953587 & -117.116756 & -115.732921 & 10.27 & 11.33 \\
 32D & $C_1$ & 116 & 0.260 & & -125.950348 & -117.105640 & -115.748493 & 11.33 & 11.40 \\
 32E & $C_s$ & 116 & 0.246 & -188.807956[1] & -125.644966 & -117.086755 & {\bf -115.767561} & 11.40 & 20.90 \\
 32F & $C_s$ & 115 & 0.011 & & -125.126341 &  -116.415892 & -115.246208 & 20.90 & \\
 33A & $C_s$ & 137 & & {\bf -204.208737} & -131.704206[1] & & & & 5.35 \\
 33B & $C_{5v}$ & 137 & & -203.575130 & -131.773513 & & & 5.35 & 5.74 \\
 33C & $C_s$ & 124 & 3.832 & & {\bf -132.287431} & -121.352047 & & 5.74 & 8.46 \\
 33D & $C_{2v}$ & 121 & 0.313 & & -131.555811 & -122.160387 & -120.665738 & 8.46 & 9.71 \\
 33E & $C_{2v}$ & 121 & 0.269 & & -131.378662 & {\bf -122.167479} & {\bf -120.741345} & 9.71 & 19.88 \\
 33F & $C_s$ & 120 & 0.011 & & -130.466899 & -121.454645 & -120.252542 & 19.88 & \\
 34A & $D_{5h}$ & 143 & & {\bf -214.068392} & -136.468311 & & & & 5.87 \\
 34B & $C_{2v}$ & 128 & 3.832 & -209.946801 & {\bf -136.797544} & -125.419532 & & 5.87 & 8.52 \\
 34C & $C_s$ & 125 & 0.313 & & -135.989024 & -126.211991 & -124.674640 & 8.52 & 9.55 \\
 34D & $C_1$ & 125 & 0.289 & & -135.963257 & {\bf -126.217685} & -124.717107 & 9.55 & 10.72 \\   
 34E & $C_s$ & 125 & 0.269 & -206.153688[2] & -135.656902 & -126.205973 & {\bf -124.748271} & 10.72 & 19.86 \\
 34F & $C_{2v}$ & 124 & 0.013 & & -135.532334 & -125.625477 & -124.283737 & 19.86 & \\
 35A & $C_s$ & 147 & & {\bf -221.771452} & -140.503355 & & & & 4.05 \\
 35B & $C_{3v}$ & 147 & & -221.293580 & -141.106305 & & & 4.05 & 5.62 \\
 35C & $C_1$ & 133 & 3.975 & & {\bf -141.957188} & -130.282153 & -126.119817 & 5.62 & 8.08 \\
 35D & $C_{2v}$ & 130 & 0.304 & & -141.402997 & {\bf -131.274427} & -129.699893 & 8.08 & 11.46 \\
 35E & $C_{2v}$ & 130 & 0.279 & -215.495670[2] & -141.051852 & -131.239157 & {\bf -129.737360} & 11.46 & 19.59 \\
 35F & $C_{3v}$ & 129 & 0.011 & & -140.132124 & -130.533259 & -129.265442 & 19.59 & \\
 36A & $C_s$ & 152 & & {\bf -230.508264} & -144.827464 & & & & 4.68 \\
 36B & $C_1$ & 152 & & -229.689435 & -145.273702 & & & 4.68 & 5.07 \\
 36C & $C_s$ & 138 & 4.351 & & {\bf -147.381965} & -134.989116 & & 5.07 & 9.51 \\
 36D & $C_1$ & 134 & 0.334 & & -145.984498 & -135.312738 & -133.656022 & 9.51 & 9.86 \\
 36E & $C_s$ & 134 & 0.303 & & -145.732293 & {\bf -135.315392} & -133.706961 & 9.86 & 11.41 \\
 36F & $C_s$ & 134 & 0.280 & -222.900558[1] & -145.378315 & -135.281551 & {\bf -133.744666} & 11.41 & 19.56 \\
 36G & $C_s$ & 133 & 0.015 & & -145.649065 & -134.766521 & -133.308212 & 19.56 & \\
 37A & $C_s$ & 158 & & {\bf -240.008130} & -149.765993 & & & & 5.25 \\
 37B & $C_1$ & 142 & 4.352 & & {\bf -151.891203} & -139.055858 & -134.764790 & 5.25 & 7.62 \\
 37C & $C_{2v}$ & 139 & 0.357 & & -151.435570 & {\bf -140.360228} & -138.622227 & 7.62 & 10.17 \\ 
 37D & $C_{2v}$ & 139 & 0.306 & -233.080175[3] & -151.061897 & -140.354330 & {\bf -138.708582} & 10.17 & 18.49 \\
 37E & $C_{3v}$ & 138 & 0.015 & & -151.100849 & -139.837570 & -138.320817 & 18.49 \\ 
 38A & $D_2$ & 164 & 30.138 & {\bf -249.159174} & -153.208710[3] & -134.319674[3] & & & 3.15 \\
 38B & $C_1$ & 163 & 27.962 &  -249.087740 & -154.165069 & -135.519468 & -129.339213 & 3.15 & 4.70 \\ 
 38C & $C_s$ & 163 & & -248.600369 & -154.041575 & & & 4.70 & 4.76 \\
 38D & $O_h$ & 144 & 0.013 & -246.414723[4] & -157.406902 & {\bf -145.849817} & {\bf -144.321054} & 4.76 & 5.40 \\
 & & & & & & & & 6.95 & \\
 38E & $C_{5v}$ & 147 & 3.714 & & {\bf -157.477108} & -144.756555 & & 5.40 & 6.95 \\
 39A & $C_{2v}$ & 169 & & {\bf -258.944962} & -158.266473 & & & & 4.48 \\
 39B & $C_{5v}$ & 153 & 5.039 & & {\bf -163.481990} & -149.455758 & & 4.48 & 9.46 \\
 39C & $C_{4v}$ & 148 & 0.014 & & -161.701728 & {\bf -149.887173} & {\bf -148.327400} & 9.46 & \\
 40A & $C_s$ & 174 & & {\bf -268.394773} & -163.956596 & & & & 4.51 \\
 40B & $C_s$ & 157 & 5.040 & -263.925318 & {\bf -167.993097} & -153.524033 & -148.492778[1] & 4.51 & 9.51 \\
 40C & $D_{4h}$ & 152 & 0.014 & -261.115599[3] & -165.996196 & {\bf -153.924517} & {\bf -152.333745} & 9.51 & \\
 41A & $C_{2v}$ & 180 & & {\bf -278.405573} & & & & & 4.60 \\
 41B & $C_s$ & 161 & 5.040 & -273.699508[1] & {\bf -172.526828} & -157.593200 & & 4.60 & 8.75 \\
\end{tabular}
\end{table}
\addtocounter{table}{-1}
\begin{table}
\caption{continued}
\vglue 2mm
\begin{tabular}{cccrccccrr}
 &  PG & $n_{nn}$ & $E_{\rm strain}$ & $\rho_0=3$ & $\rho_0=6$ & $\rho_0=10$ \ & 
$\rho_0=14$ \ & $\rho_{min}$ & $\rho_{max}$ \\
\hline
 41C & $C_{2v}$ & 157 & 0.435 & & -171.124667 & -158.505557 & -156.511959 & 8.75 & 9.60 \\
 41D & $C_s$ & 157 & 0.368 & -269.034859[4] & -170.808262 & {\bf -158.520688} & -156.630647 & 9.60 & 11.06 \\
 41E & $C_{2v}$ & 157 & 0.366 & -267.688391[4] & -170.758682 & -158.518482 & {\bf -156.633479} & 11.06 & 16.83 \\
 41F & $C_{s}$ & 156 & 0.017 & & -170.902496 & -158.066916 & -156.359354 & 16.83 & \\
 42A & $C_{3v}$ & 186 & & {\bf -288.335415} & -172.252388[1] & & & & 4.54 \\
 42B & $C_s$ & 166 & 5.275 & -282.296210[1] & {\bf -177.680222} & -162.365546 & -157.124535 & 4.54  & 9.77 \\
 42C & $C_s$ & 161 & 0.368 & & -175.184389 & {\bf -162.565848} & -160.638215 & 9.77 & 11.10 \\
 42D & $C_s$ & 161 & 0.367 & & -175.135897 & -162.563540 & {\bf -160.641020} & 11.10 & 16.82 \\
 42E & $C_s$ & 160 & 0.017 & & -175.405455 & -162.133633 & -160.371244 & 16.82 & \\
 43A & $C_{2v}$ & 192 & & {\bf -298.172449} & -175.540353[2] & & & & 4.40 \\
 43B & $C_s$ & 171 & & -290.632163[2] & {\bf -183.092699} & & & 4.40 & 9.34 \\
 43C & $C_{2v}$ & 166 & 0.443 & & -180.837781 & -167.585928 & -165.511347 & 9.34 & 9.57 \\
 43D & $C_{2v}$ & 166 & 0.374 & -288.301568[2] & -180.515098 & {\bf -167.602632} & {\bf -165.634973} & 9.57 & 16.66 \\
 43E & $C_1$ & 165 & 0.017 & & -180.662511 & -167.152124 & -165.372773 & 16.66 & \\
 44A & $C_{5v}$ & 198 & & {\bf -308.277011} & -179.032310[8] & & & & 4.38 \\
 44B & $C_1$ & 175 & 5.722 & -302.003640 & {\bf -187.626292} & -171.071131 & -165.689737 & 4.38 & 9.38 \\
 44C & $C_s$ & 170 & 0.443 & & -185.214201 & -171.631403 & -169.519034 & 9.38 & 9.65 \\
 44D & $C_s$ & 170 & 0.374 & -300.113694[3] & -184.841212 & {\bf -171.645465} & {\bf -169.642441} & 9.65 & 16.65 \\
 44E & $C_1$ & 169 & 0.020 & & -185.296321 & -171.226450 & -169.385406 & 16.65 \\
 45A & $I_h$ & 204 & & {\bf -318.660653} & -182.301077[24] & & & & 4.24 \\
 45B & $C_1$ & 180 & 5.965 & -310.750385 & {\bf -192.954739} & -175.389424 & -169.712668 & 4.24 & 8.72 \\
 45C & $C_s$ & 175 & 0.691 & & -191.230983 & -176.548074 & -174.095718 & 8.72 & 8.95 \\
 45D & $C_{2v}$ & 175 & 0.452 & & -190.548331 & {\bf -176.665997} & -174.510080 & 8.95 & 11.79 \\
 45E & $C_{2v}$ & 175 & 0.451 & -303.604189[3] & -190.500098 & -176.663202 & {\bf -174.511633} & 11.79 & 15.03 \\
 45F & $C_s$ & 174 & 0.018 & &  -190.708793 & -176.296327 & -174.397893 & 15.03 & \\
 46A & $C_s$ & 207 & & {\bf -327.033118} & & & & & 3.96 \\
 46B & $C_{2v}$ & 186 & 6.646 & -320.118738[1] &  {\bf -199.177751} & {\bf -181.236182} & -174.605103[1] & 3.96 & 10.46 \\
 46C & $C_s$ & 179 & 0.452 & & -194.923850 & -180.711434 & -178.517769 & 10.46 & 11.79 \\
 46D & $C_s$ & 179 & 0.451 & & -194.876995 & -180.708654 & {\bf -178.519320} & 11.79 & 14.97 \\ 
 46E & $C_s$ & 178 & 0.019 & & -195.426416 & -180.401063 & -178.416567 & 14.97 & \\
  47A & $C_3$ & 210 & 40.212 & {\bf -336.666189} & -194.242255 & -170.368382 & & & 4.03 \\
  47B & $C_1$ & 190 & 6.648 & -331.591008 & {\bf -203.704178} & -185.301500 & -178.621197 & 4.03 & 9.60 \\ 
  47C & $C_{2v}$ & 184 & 0.461 & -323.010894[6] & -200.256505 & {\bf -185.745782} & {\bf -183.508227} & 9.60 & 14.86 \\
  47D & $C_s$ & 183 & 0.019 & & -200.473614 & -185.381591 & -183.411312 & 14.86 & \\
  48A & $C_{2v}$ & 213 & & {\bf -346.662788} & & & & & 4.01 \\
  48B & $C_s$ & 195 & 7.464 & -340.395646[1] & {\bf -209.044000} & -189.5566723 & -182.617917[1] & 4.01 & 8.01 \\
  48C & $C_{2v}$ & 190 & 0.808 & & -207.541529 & {\bf -191.586299} & {\bf -188.888965} & 8.01 & 20.22 \\
  48D & $C_s$ & 187 & 0.020 & & -205.418604 & -189.535478 & -187.440413 & 20.22 & \\
  49A & $C_s$ & 219 & & {\bf -356.412817} & & & & & 3.91 \\
  49B & $C_{3v}$ & 201 & 7.770 & -350.540378[1] & {\bf -215.253702} & -195.275059 & -187.212009 & 
3.91 & 9.70 \\
  49C & $C_s$ & 194 & 0.808 & & -211.978083 & {\bf -195.639320} & {\bf -192.898412} & 9.70 & 16.14 \\
  49D & $C_{3v}$ & 192 & 0.021 & & -211.117300 & -194.627748 & -192.455390 & 16.14 & \\
  50A & $C_{5v}$ & 224 & & {\bf -366.635589} & & & & & 4.04 \\
  50B & $C_s$ & 205 & 7.770 & & {\bf -219.820229} & -199.344795 & -191.239955 & 4.04 & 8.85 \\
  50C & $C_s$ & 199 & 0.809 & -360.600022[2] & -217.406110 & {\bf -200.691615} & -197.906988 & 8.85 & 10.50 \\
  50D & $D_{3h}$ & 198 & 0.020 & & -217.432647 & -200.640183 & {\bf -198.455633} & 10.50 & \\
  51A & $C_2$ & 232 & & {\bf -376.673413} & -217.627962 & & & & 3.74 \\
  51B & $C_{2v}$ & 210 & 9.360 & -373.932555[1] & {\bf -225.391240} & -202.911593 & & 3.74 & 6.59 \\
  51C & $C_s$ & 210 & 8.551 & -374.031668[1] & -225.236760 & -203.643178 & -195.286917 & 6.59 & 9.18 \\
  51D & $C_s$ & 203 & 0.809 & & -221.842387 & {\bf -204.744598} & -201.916433 & 9.18 & 10.33 \\
  51E & $C_s$ & 202 & 0.023 & & -222.080140 & -204.714809 & {\bf -202.468274} & 10.33 & \\
  52A & $C_s$ & 238 & & {\bf -387.587332 } & & & & & 3.71 \\
  52B & $C_{3v}$ & 216 & 9.618 & -384.520749 & {\bf -231.615013} & -208.673261 & -199.353198 & 3.71 & 6.61 \\
  52C & $C_{2v}$ & 216 & 8.840 & & -231.443769 & -209.379230 & -199.715952 & 6.61 & 9.72 \\
  52D & $C_{2v}$ & 208 & 0.809 & -380.031663[3] & -227.268275 & {\bf -209.796728} & -206.924957 & 9.72 & 10.15 \\
  52E & $C_{2v}$ & 207 & 0.022 & & -227.500884 & -209.784832 & {\bf -207.480764} & 10.15 & \\
\end{tabular}
\end{table}
\addtocounter{table}{-1}
\begin{table}
\caption{continued}
\vglue 2mm
\begin{tabular}{cccrccccrr}
 &  PG & $n_{nn}$ & $E_{\rm strain}$ & $\rho_0=3$ & $\rho_0=6$ & $\rho_0=10$ \ 
& $\rho_0=14$ \ & $\rho_{min}$ & $\rho_{max}$ \\
\hline
 53A & $D_{6d}$ & 244 & & {\bf -398.783184} & & & & & 3.70 \\
 53B & $C_{2v}$ & 222 & 9.902 & -395.150063 & {\bf -237.834976} & {\bf -214.409721} & -203.180586[1] & 3.70 & 10.30 \\
 53C & $C_{s}$ & 211 & 0.025 & & -232.147467 & -213.859451 & {\bf -211.493405} & 10.30 & \\
 54A & $C_s$ & 248 & & {\bf -407.966010} & -232.987285 & & & & 3.41 \\
 54B & $C_{5v}$ & 228 & 10.215 & -405.831836 & {\bf -244.058174} & {\bf -220.118611} & -208.329991 & 3.41 & 10.32 \\
 54C & $C_{2v}$ & 217 & 0.465 & & -238.114592 & -219.426691 & {\bf -216.636864} & 10.32 & 15.10 \\
 54D & $C_{2v}$ & 216 & 0.024 & & -237.568575 & -218.929453 & -216.505895 & 15.10 & \\
 55A & $C_1$ & 252 & & {\bf -417.918562} & & & & & 3.25 \\
 55B & $I_h$ & 234 & 10.543 & -416.625645 & {\bf -250.286609} & {\bf -225.814286} & -213.523774 & 3.25 & 11.15 \\
 55C & $C_{2v}$ & 221 & 0.465 & & -242.622450 & -223.482018 & {\bf -220.646208} & 11.15 & 15.09 \\
 55D & $C_s$ & 220 & 0.027 & & -242.211381 & -223.003991 & -220.518533 & 15.09 & \\
 56A & $C_s$ & 258 & & {\bf -428.611289} & & & & & 3.60 \\
 56B & $C_{3v}$ & 237 & 10.545 & -425.709433 & {\bf -253.922955} & {\bf -228.900154} & -216.537609 
& 3.60 & 10.18 \\
 56C & $C_{2v}$ & 226 & 0.481 & & -248.167730 & -228.552734 & -225.633470 & 10.18 & 11.38 \\
 56D & $C_s$ & 226 & 0.465 & & -248.051289 & -228.533805 & {\bf-225.655136} & 11.38 & 15.07 \\
 56E & $D_{3h}$ & 225 & 0.026 & & -247.635625 & -228.074040 & -225.531024 & 15.07 & \\
 57A & $D_{3h}$ & 264 & & {\bf -439.960320} & & & & & 4.13 \\
 57B & $C_s$ & 241 & 10.563 & -436.124573[1] & {\bf -258.041717} & -232.877539 & -220.502856 & 4.13 & 9.61 \\
 57C & $C_{2v}$ & 231 & 0.465 & -426.761701[3] & -253.482114 & {\bf -233.585699} & {\bf -230.663986} & 
9.61 & 21.83 \\
 57D & $C_s$ & 229 & 0.028 & & -252.749395 & -232.226794 & -229.557450 & 21.83 & \\
 58A & $C_s$ & 267 & & {\bf -449.432282} & -253.758845[1] & & & & 4.05 \\
 58B & $C_{3v}$ & 246 & 10.585 & -442.501104[1] & {\bf -263.410755} & -237.898423 & -225.469489 & 
4.05 & 9.53 \\
 58C & $D_{3h}$ & 237 & 1.467 & & -260.086805 & {\bf -238.658866} & {\bf -234.809078} & 9.53 & 14.63 \\
 58D & $C_{1}$ & 235 & 0.465 & & -257.876211 & -237.631296 & -234.671473 & 14.63 & 14.94 \\
 58E & $C_{3v}$ & 234 & 0.027 & & -258.241372 & -237.298283 & -234.569986 & 14.94 & \\
 59A & $C_{3v}$ & 267 & & {\bf -459.509280} & -260.795085[1] & & & & 3.77 \\
 59B & $C_{2v}$ & 272 & & -459.250400 & -261.475383[2] & & & 3.77 & 4.26 \\
 59C & $C_1$ & 250 & 10.832 & & -267.857802 & -241.705877 & -229.053528 & 4.26 & 4.86 \\
     & & & & & & & & 7.93 & 9.09 \\
 59D & $C_{2v}$ & 250 & 10.939 & -453.109891[2] & {\bf -267.945226} & -241.627254 & -228.849111 & 
4.86 & 7.93 \\
 59E & $T_d$ & 240 & 0.027 & & -264.710286  & {\bf -243.330508} & {\bf -240.572493} & 9.09 & \\
 60A & $C_s$ & 278 & & {\bf -470.448485} & & & & & 4.51 \\
 60B & $C_s$ & 255 & 11.042 & & {\bf -273.341243} & -246.579281 & -233.702420 & 4.51 & 9.29 \\
 60C & $C_{2v}$ & 246 & 1.470 & & -269.958055 & -247.763712 & -243.823553 & 9.29 & 9.85 \\
 60D & $C_s$ & 245 & 0.540 & & -269.408035 & {\bf -247.785988} & -244.566234 & 9.85 & 11.00 \\
 60E & $C_{2v}$ & 245 & 0.536 & & -269.292014 & -247.781611 & -244.572017 & 11.00 & 13.95 \\
 60F & $C_s$ & 244 & 0.027 & & -269.050233 & -247.370671 & {\bf -244.579066} & 13.95 & \\
 61A & $T_d$ & 284 & & {\bf -482.025765} & -271.692826 & & & & 4.65 \\
 61B & $C_{2v}$ & 260 & 11.174 & & {\bf -278.726626} & -251.502374 & -238.498558 & 4.65 & 9.54 \\ 
 61C & $C_{3v}$ & 249 & 0.027 & & -274.114496 & {\bf -252.402685} & {\bf -249.587740} & 9.54 & \\
 62A & $C_s$ & 293 & & {\bf -491.052378} & -272.267523 & & & & 4.00 \\
 62B & $C_s$ & 264 & 11.707 & -485.425587[3] & {\bf -283.183002} & -255.038209 & -241.549494 & 4.00 & 6.33 \\ 
 62C & $C_{2v}$ & 264 & 11.174 & & -283.102592 & -255.558891 & -242.509563 & 6.33 & 9.22 \\
 62D & $C_{2v}$ & 255 & 1.473 & & -279.826688 & -256.868159 & -252.837765 & 9.22 & 9.64 \\
 62E & $C_1$ & 254 & 0.623 & & -279.851878 & -256.899841 & -253.442114 & 9.64 & 9.78 \\
 62F & $C_s$ & 254 & 0.560 & & -279.460422 & {\bf -256.908198} & -253.554702 & 9.78 & 10.47 \\
 62G & $C_1$ & 254 & 0.526 & & -279.357773 & -256.897318 & {\bf -253.612942} & 10.47 & 14.14 \\
 62H & $C_1$ & 253 & 0.027 & & -278.499195 & -256.445284 & -253.594448 & 14.14 & \\
 63A & $C_s$ & 301 & & {\bf -501.731893} & & & & & 3.83 \\
 63B & $C_1$ & 269 & 11.880 & -497.181562 & {\bf -288.560948} & -259.921304 & -246.279386 & 3.83 & 8.87 \\
 63C & $C_s$ & 259 & 0.533 & 495.356854[1] & -285.141958 & {\bf-262.030007} & {\bf -258.620607} & 8.87 & 14.14 \\
 63D & $C_s$ & 258 & 0.027 & & -283.876854 & -261.487847 & -258.601156 & 14.14 & \\
 64A & $D_{3d}$ & 307 & & {\bf -512.831683} & & & & & 3.85 \\
 64B & $C_s$ & 274 & 12.066 & -503.234962 & {\bf -293.931716} &  -264.790492 & -250.989419 & 3.85 & 8.22 \\
 64C & $C_{2v}$ & 265 & 0.552 & -501.818445[4] & -291.412250 & {\bf -268.023828} & {\bf -264.587042} & 8.22 & 24.31 \\
 64D & $C_s$ & 262 & 0.031 & & -289.333587 & -265.722635 & -262.644509 & 24.31 & \\
\end{tabular}
\end{table}
\addtocounter{table}{-1}
\begin{table}
\caption{continued}
\vglue 2mm
\begin{tabular}{cccrccccrr}
 &  PG & $n_{nn}$ & $E_{\rm strain}$ & $\rho_0=3$ & $\rho_0=6$ & $\rho_0=10$ \ & 
$\rho_0=14$ \ & $\rho_{min}$ & $\rho_{max}$ \\
\hline
 65A & $C_s$ & 310 & & {\bf -523.446427} & & & & & 3.86 \\
 65B & $C_1$ & 278 & 12.494 & & -298.321863 & -268.406866 & -254.407364 & 3.86 & 4.74 \\
 65C & $C_2$ & 278 & 12.574 & -515.414573 & {\bf -298.392345} & -268.351634 & -254.301092 & 4.74 & 6.32 \\
 65D & $C_1$ & 278 & 12.081 & & -298.317501 & -268.836726 & -254.984242 & 6.32 & 8.12 \\
 65E & $C_{2v}$ & 269 & 0.603 & & -296.317862 & {\bf -272.098786} & -268.513581 & 8.12 & 10.53 \\
 65F & $C_1$ & 269 & 0.565 & & -296.027614  & -272.083782 & -268.578320 & 10.53 & 11.30 \\
 65G & $C_1$ & 269 & 0.552 & & -295.806729 & -272.069820 & {\bf -268.594702} & 11.30 & 19.71 \\
 65H & $C_s$ & 267 & 0.027 & & -293.637761 & -270.572952 & -267.614569 & 19.71 & \\
 66A & $C_s$ & 314 & & {\bf -534.464040} & & & & & 3.82 \\
 66B & $C_1$ & 284 & 15.128 & -530.516639 & -303.512725 & -271.712472 & -256.391481 & 3.82 & 4.83 \\
 66C & $C_1$ & 283 & 12.758 & -528.936842 & {\bf -303.763297} & -273.223079 & -259.078765 & 4.83 & 7.83 \\
 66D & $C_s$ & 274 & 0.640 & & -301.866029 & -277.134660 & -273.454381 & 7.83 & 9.74 \\
 66E & $C_s$ & 274 & 0.573 & & -301.464889 & {\bf -277.145223} & -273.573502 & 9.74 & 11.50 \\
 66F & $C_s$ & 274 & 0.552 & & -301.193166 & -277.115736 & {\bf -273.602343} & 11.50 & 24.31 \\
 66G & $C_s$ & 271 & 0.033 & & -299.437330 & -274.866840 & -271.669464 & 24.31 & \\
 67A & $C_s$ & 318 & & {\bf -544.754257} & -297.547886 & & & & 3.80 \\
 67B & $C_s$ & 289 & 15.345 & -539.666842[2] & -308.979846 & -276.564507 & & 3.80 & 5.46 \\
 67C & $C_2$ & 288 & 12.945 & -538.179357[1] & {\bf -309.130322} & -278.090470 & -263.853175 & 5.46 & 7.72\\
 67D & $C_{2v}$ & 279 & 0.677 & & -307.349356 & {\bf -282.171568} & -278.396059 & 7.72 & 11.36 \\
 67E & $C_{2v}$ & 279 & 0.674 & & -307.234099 & -282.166219 & {\bf -278.400953} & 11.36 & 21.85 \\
 67F & $C_{3v}$ & 276 & 0.033 & & -304.922619 & -279.938518 & -276.682086 & 21.85 & \\
 68A & $C_s$ & 324 & & {\bf -555.582496} & & & & & 3.68 \\
 68B & $C_1$ & 294 & 15.980 & -551.451174 & {\bf -314.374880} & -281.011142 & -265.753769 & 3.68 & 7.29 \\
 68C & $C_s$ & 293 & 12.938 & -549.985909[1] & -314.146641 & -281.905282 & -267.100885 & 7.29 & 7.78 \\
 68D & $C_1$ & 283 & 0.666 & & -311.913082 & -286.249826 & -282.429781 & 7.78 & 9.67 \\
 68E & $C_{2v}$ & 283 & 0.597 & & -311.515987 & {\bf -286.263045} & -282.554370 & 9.67 & 11.24 \\
 68F & $C_s$ & 283 & 0.573 & & -311.246697 & -286.237292 & -282.588836 & 11.24 & 11.51 \\
 68G & $C_{2v}$ & 283 & 0.552 & & -310.973538 & -286.207609 & -282.617642 & 11.51 & 13.46 \\
 68H & $C_{3v}$ & 282 & 0.032 & & -311.238864 & -285.954388 & {\bf-282.683003} & 13.46 & \\
 69A & $C_1$ & 329 & & {\bf -566.364140} & -307.245602 & & & & 3.65 \\
 69B & $C_1$ & 308 & & -560.639499 & -318.205833 & & & 3.65 & 4.22 \\
 69C & $C_{5v}$ & 298 & 14.909 & -559.783221[1] & -319.687802 & -286.265131 & & 4.22 & 5.22 \\
 69D & $C_s$ & 299 & 16.223 & & {\bf -319.819905} & -285.836885[1] & & 5.22 & 7.64 \\
 69E & $C_s$ & 288 & 0.705 & & -317.396150 & -291.284253 & -287.366805 & 7.64 & 9.66 \\
 69F & $C_1$ & 288 & 0.653 & & -317.293147 & {\bf -291.294050} & {\bf -287.462110} & 9.66 & 18.06 \\
 69G & $C_1$ & 286 & 0.034 & & -315.861945 & -290.031196 & -286.695880 & 18.06 & \\
 70A & $C_{2v}$ & 335 & & {\bf -577.739782} & & & & & 3.22 \\
 70B & $C_1$ & 332 & & -577.286914 & -313.362561 & & & 3.22 & 3.67 \\
 70C & $C_5$ & 314 & & -571.815209 & -324.211871 & & & 3.67 & 4.16 \\
 70D & $C_{5v}$ & 304 & 15.324 & -570.747100 & {\bf -325.887749} & -291.872039 & & 4.16 & 7.89 \\
 70E & $C_s$ & 293 & 0.676 & & -323.082118 & {\bf -296.412149} & -292.439398 & 7.89 & 12.18 \\
 70F & $C_1$ & 293 & 0.659 & & -322.802226 & -296.365659 & {\bf -292.462856} & 12.18 & 17.96 \\
 70G & $C_s$ & 291 & 0.035 & & -321.499078 & -295.108247 & -291.708765 & 17.96 & \\
 71A & $C_1$ & 330 & & {\bf -588.396687} & -316.491227 & & & & 3.42 \\
 71B & $C_5$ & 320 & & -583.202864 & -330.363241 & & & 3.42 & 4.28 \\
 71C & $C_{5v}$ & 310 & 17.608 & -581.944905[1] & {\bf -331.588748} & -295.611936[2] & & 4.28 & 7.29 \\
 71D & $C_{2v}$ & 299 & 0.695 & & -329.356130 & {\bf -302.405357} & {\bf -298.405353} & 7.29 & 21.50 \\
 71E & $C_s$ & 296 & 0.034 & & -326.944961 & -300.178953 & -296.721356 & 21.50 \\
 72A & $C_1$ & 335 & & {\bf -599.690310} & -320.666142 & & & & 3.40 \\
 72B & $C_1$ & 314 & 17.960 & -595.064909 & -335.877119 & -299.300783 & -283.195227 & 3.40 & 4.77 \\
 72C & $C_s$ & 313 & 16.818 & -592.641021 & {\bf -336.121753} & -299.750026 & -283.664026 & 4.77 & 7.29 \\
 72D & $C_1$ & 303 & 0.713 & & -333.967611 & {\bf -306.460206} & -302.387256 & 7.29 & 10.53 \\
 72E & $C_1$ & 303 & 0.695 & & -333.804475 & -306.453999 & {\bf -302.413229} & 10.53 & 21.49 \\
 72F & $C_s$ & 300 & 0.038 & & -331.710111 & -304.261423 & -300.734505 & 21.49 & \\
\end{tabular}
\end{table}
\addtocounter{table}{-1}
\begin{table}
\caption{continued.}
\vglue 2mm
\begin{tabular}{cccrccccrr}
 &  PG & $n_{nn}$ & $E_{\rm strain}$ & $\rho_0=3$ & $\rho_0=6$ & $\rho_0=10$ \ & 
$\rho_0=14$ \ & $\rho_{min}$ & $\rho_{max}$ \\
\hline
 73A & $C_3$ & 339 & & {\bf -610.936684} & -324.743483 & & & & 3.50 \\
 73B & $C_s$ & 319 & 18.316 & -603.873759[2] & {\bf -341.266253} & -304.026668 & -287.806170 & 3.50 & 6.97 \\
 73C & $C_s$ & 318 & 16.192 & -602.726569[2] & -340.979140 & -305.189678 & -289.451450 & 6.97 & 7.03 \\
 73D & $C_s$ & 308 & 0.725 & & -339.404682 & {\bf -311.517522} & -307.374386 & 7.03 & 10.72 \\
 73E & $C_s$ & 308 & 0.695 & & -339.241592 & -311.502535 & {\bf -307.421094} & 10.72 & 21.49 \\
 73F & $C_1$ & 305 & 0.037 & & -337.195061 & -309.332640 & -305.747113 & 21.49 & \\
 74A & $C_{3v}$ & 357 & & {\bf -622.679870} & & & & & 3.92 \\
 74B & $C_s$ & 324 & 18.513 & -614.020203[2] & {\bf -346.610834} & -308.886442 & & 3.92 & 6.83 \\ 
 74C & $C_{5v}$ & 313 & 0.699 & -610.492247[3] & -345.199617 &  {\bf -316.650760} & {\bf -312.441302} & 6.83 & 21.38 \\
 74D & $C_s$ & 310 & 0.036 & & -342.649207 & -314.403461 & -310.759707 & 21.38 \\
 75A & $C_s$ & 364 & & {\bf-633.513370} & & & & & 3.70 \\
 75B & $C_1$ & 328 & 18.559 & -630.521082 & -351.177041 & -312.987148 & -296.598914 & 3.70 & 5.81 \\
 75C & $D_{5h}$ & 319 & 0.718 & & {\bf -351.472365} & {\bf -322.643558} & {\bf -318.407330} & 5.81 & 21.14 \\
 75D & $C_s$ & 316 & 0.035 & & -348.965031 & -320.414355 & -316.759660 & 21.14 & \\
 76A & $C_s$ & 370 & & {\bf -644.951602} & & & & & 3.56 \\
 76B & $C_1$ & 334 & 21.847 & -642.129130 & -355.970657 & -315.525026 & -298.048750 & 3.56 & 4.16 \\
 76C & $C_1$ & 333 & 19.297 & -641.559041 & -356.249587 & -317.224994 & -300.488549 & 4.16 & 4.40 \\
 76D & $C_1$ & 333 & 19.272 & -639.714777 & {\bf -356.372708} & -317.247110 & -300.338642 & 4.40 & 6.17 \\
 76E & $C_s$ & 323 & 0.737 & & -356.085310 & {\bf -326.697484} & -322.386901 & 6.17 & 10.93 \\
 76F & $C_{2v}$ & 323 & 0.718 & & -355.771608 & -326.682964 & {\bf -322.414257} & 10.93 & 21.14 \\
 76G & $C_1$ & 320 & 0.035 & & -353.408271 & -324.459592 & -320.766507 & 21.14 & \\
 77A & $C_1$ & 368 & & {\bf -656.079789} & -349.695795 & & & & 3.68 \\
 77B & $C_s$ & 339 & 22.562 & -650.115580[2] & -361.292679 & -319.897701[2] & & 3.68 & 4.97 \\
 77C & $C_1$ & 338 & 19.477 & & {\bf -361.727086} & -322.095213 & -305.018458 & 4.97 & 6.12 \\
 77D & $C_{2v}$ & 328 & 0.749 & & -361.520971 & {\bf -331.753953} & {\bf -327.371999} & 6.12 & 20.66 \\
 77E & $C_s$ & 325 & 0.035 & & -358.839797 & -329.504700 & -325.773351 & 20.66 & \\
 78A & $C_3$ & 384 & & {\bf -667.576295} & & & & & 3.54 \\
 78B & $C_1$ & 344 & 22.724 & -664.005603 & -366.727436 & -324.814567 & -307.287362 & 3.54 & 5.18 \\
 78C & $C_1$ & 344 & 22.651 & & {\bf -366.761455} & -324.883480 & -307.076218 & 5.18 & 6.04 \\
 78D & $C_s$ & 338 & 11.849 & & -366.722670 & -330.163464 & -315.401197 & 6.04 & 6.53 \\
 78E & $C_1$ & 332 & 0.769 & & -366.132260 & {\bf -335.806666} & -331.349274 & 6.53 & 10.71 \\
 78F & $C_s$ & 332 & 0.749 & -650.435529[3] & -365.871754 & -335.796094 & {\bf -331.379143} & 10.71 & 16.79 \\
 78G & $C_1$ & 330 & 0.037 & & -364.447522 & -334.579848 & -330.786061 & 16.79 & \\
 79A & $C_1$ & 385 & & {\bf -678.940231} & -358.579695 & & & & 3.59 \\ 
 79B & $C_{2v}$ & 348 & 19.381 & -673.564685 & {\bf -372.832290} & -332.365043 &  & 3.59  & 6.53 \\
 79C & $C_{2v}$ & 343 & 11.973 & & -372.465769 & -335.139012 & -320.222688 & 6.53 & 6.67 \\
 79D & $C_{2v}$ & 337 & 0.783 & -663.444178 & -371.568226 & {\bf -340.862137} & -336.332369 & 6.67 & 11.06 \\
 79E & $C_{s}$ & 337 & 0.752 & -659.185953[3] & -371.199420 & -340.834084 & -336.380124 & 11.06 & 11.48 \\
 79F & $D_{3h}$ & 336 & 0.036 & -661.623659[2] & -370.982315 & -340.653542 & {\bf -336.798725} & 11.48 & \\
 80A & $C_3$ & 387 & & {\bf -690.577890} & & & & & 3.53 \\
 80B & $C_s$ & 354 & 21.858 & -683.220844[2] & {\bf -378.333471} & -335.897891[1] & & 3.53 & 6.36 \\
 80C & $C_1$ & 348 & 12.120 & & -377.972497 & -340.072278 & -324.944256 & 6.36 & 7.34 \\
 80D & $C_1$ & 341 & 0.806 & -674.292236 & -376.175628 & {\bf -344.911381} & -340.304108 & 7.34 & 10.44 \\
 80E & $C_s$ & 341 & 0.783 & & -375.919374 & -344.904384 & -340.339544 & 10.44 & 11.02 \\
 80F & $C_s$ & 341 & 0.752 & -668.581802[4] & -375.572469 & -344.877983 & -340.387336 & 11.02 & 11.36 \\
 80G & $C_s$ & 340 & 0.040 & & -375.635945 & -344.728294 & {\bf -340.811371} & 11.36 & \\
\end{tabular}
\end{table}
\begin{multicols}{2}

\end{multicols}

\begin{thebibliography}{100}
\bibitem{JD95cnote}
In ref.~\onlinecite{JD95c}, $\gamma(E')_s$ incorrectly appeared in the equation
  for $p_i(E)$, the probability that the cluster resides in region $i$ of
  configuration space.
\end{thebibliography}
\end{document}